\documentclass[]{aa}  

\usepackage{graphicx}
\usepackage[varg]{txfonts}
\usepackage[]{hyperref}
\usepackage[percent]{overpic}
\usepackage{natbib}
\usepackage{lscape}
\usepackage{subfig}
\usepackage{subfloat}
\usepackage{amssymb}
\usepackage{enumitem}

\begin{document} 
\renewcommand{\theenumi}{\Alph{enumi}}

\title{Nowhere to Hide: Radio-faint AGN in the GOODS-N field \thanks{a machine readable catalogue accompanies this paper}}

\subtitle{I. Initial catalogue and radio properties}

\author{J. F.~Radcliffe\inst{1,2,3}
\and M. A.~Garrett\inst{2,4} 
\and T. W. B.~Muxlow\inst{2}
\and R. J.~Beswick\inst{2}
\and P. D.~Barthel\inst{1}
\and A. T.~Deller\inst{5}
\and A.~Keimpema\inst{6}
\and R. M. ~Campbell\inst{6}
\and N. ~Wrigley\inst{2}}

\institute{Kapteyn Astronomical Institute, University of Groningen, 9747 AD Groningen, The Netherlands \\
\email{j.f.radcliffe@astro.rug.nl}
\and Jodrell Bank Centre for Astrophysics/e-MERLIN, The University of Manchester, M13 9PL, United Kingdom
\and ASTRON, the Netherlands Institute for Radio Astronomy, Postbus 2, 7990 AA Dwingeloo, The Netherlands
\and Leiden Observatory, Leiden University, PO Box 9513, 2300 RA Leiden, The Netherlands
\and Centre for Astrophysics and Supercomputing, Swinburne University of Technology, PO Box 218 Hawthorn, VIC 3122, Australia
\and Joint Institute for VLBI ERIC, Postbus 2, 7990 AA Dwingeloo, The Netherlands}

\date{Received <date> / Accepted <date>}

\titlerunning{There's nowhere to run, nowhere to hide}
\authorrunning{J. F. Radcliffe et al.}

 
  \abstract
   {The occurrence of active galactic nuclei (AGN) is critical to our understanding of galaxy evolution and formation. Radio observations provide a crucial, dust-independent tool to study the role of AGN. However, conventional radio surveys of deep fields ordinarily have arc-second scale resolutions often insufficient to reliably separate radio emission in distant galaxies originating from star-formation and AGN-related activity. Very long baseline interferometry (VLBI) can offer a solution by identifying only the most compact radio emitting regions in galaxies at cosmological distances where the high brightness temperatures (in excess of $10^5$ K) can only be reliably attributed to AGN activity.} 
   {We present the first in a series of papers exploring the faint compact radio population using a new wide-field VLBI survey of the GOODS-N field. This will expand upon previous surveys, permitting the characterisation of the faint, compact radio source population in the GOODS-N field. The unparalleled sensitivity of the European VLBI Network (EVN) will probe a luminosity range rarely seen in deep wide-field VLBI observations, thus providing insights into the role of AGN to radio luminosities of the order $10^{22}~\mathrm{W\,Hz^{-1}}$ across cosmic time.}
   {The newest VLBI techniques are used to completely cover an entire 7.\!\arcmin5 radius area to milliarcsecond resolutions, while bright radio sources ($S > 0.1$\,mJy) are targeted up to 25\arcmin~from the pointing centre. Multi-source self-calibration, and a primary beam model for the EVN array are used to correct for residual phase errors and primary beam attenuation respectively.}
   {This paper presents the largest catalogue of VLBI detected sources in GOODS-N comprising of 31 compact radio sources across a redshift range of 0.11\mbox{-}3.44, almost three times more than previous VLBI surveys in this field. We provide a machine-readable catalogue and introduce the radio properties of the detected sources using complementary data from the e-MERLIN Galaxy Evolution survey (eMERGE).}
   {}


   \maketitle
%

\section{Introduction}\label{Sec: Introduction}
Radio source counts above mJy flux densities are dominated by radio galaxies and quasars powered by active galactic nuclei (AGN). Below mJy flux densities, there is an observed upturn far in excess of those predicted by extrapolating source counts of high luminosity radio galaxies and quasars. This upturn is found to comprise an increasing fraction of active star forming galaxies and faint ‘non-jetted’ or radio-quiet AGN plus a decreasing fraction of classical radio-loud sources \citep[see][and references therein]{prandoni2001,Huynh2015,Padovani2016}. The majority of extragalactic radio surveys are carried out at arc-second resolutions (corresponding to galactic/$\sim$10's kpc physical scales at $z\geq0.1$) where it can be difficult to distinguish between the sub-kpc scale AGN activity and the kpc star-formation related emission based purely on their radio morphologies. This is particularly important if we are to characterise the properties of radio-quiet AGN whose radio emission in local systems are confined within the host galaxy \citep[see][and references therein]{Orienti:2015sk}. As a result, these surveys rely on multi-wavelength diagnostics, such as radio-excess, SED fitting, X-ray emission etc., in order to identify any AGN activity \citep[e.g.][]{Bonzini2013:cd,Smolcic:2017ef}. These diagnostics are often incomplete with dust masking the signatures of AGN activity. For example, X-rays often do not detect Compton-thick AGN which are estimated to account for over a third of the total AGN population \citep[][]{Mateos:2017fm}.

These hidden AGN can be found using high resolution, dust-independent radio observations. Indeed, surveys using e-MERLIN, such as the \textbf{e-MER}lin \textbf{G}alaxy \textbf{E}volution (e-MERGE) survey (Muxlow et al. in prep., \citet{muxlow2005high}), and Very Long Baseline Interferometry (VLBI) \citep[e.g.][]{Middelberg:2011bx,Middelberg:2013hy,Ruiz:2017ur} have shown that deep, sub-arcsecond and sub-kpc observations can effectively isolate AGN activity from compact star-forming related emission in distant galaxies.
 
VLBI observations detect bright, compact objects with brightness temperatures in excess of $10^{5}$\,K. In nearby galaxies, these brightness temperatures can be typically reached by either AGN, supernovae (SNe) and their remnants (SNRs). However, in more distant galaxies ($z > 0.1$), these brightness temperatures can typically only be attained by AGN-related emission processes \citep[e.g.][]{kewley2000compact}, thus making VLBI a unique and invaluable tool to survey distant galaxies for AGN activity. However, until the last decade, there have been many factors preventing VLBI from being used as a survey instrument.

Conventional wide-field VLBI observations mapped a significant proportion of the primary beam by using a single correlation pass at a ultra-fine temporal and frequency resolution in order to limit time and bandwidth smearing towards the edge of the primary beam \citep{garrett2001agn}. As a result, the observer would receive a single large and unwieldy (often $\sim$TB size) data set. With the ever increasing number of VLBI-ready telescopes along with widening bandwidths, the bit rates of modern VLBI arrays are rapidly increasing and this method of correlation has become computationally infeasible. Software correlators established the `multiple simultaneous phase centre observing'  approach to correlation \citep{deller2011difx,MorganVLBI2011,keimpema2015sfxc} which substantially reduces the computational load. Here, the observer defines a number of sub-fields (also referred to as phase centres) which can be either sources of interest or can be arranged to cover the entire primary beam. When correlated, these data are split and each sub-section is correlated at the ultra high temporal and frequency resolution required to restrain smearing. It is then copied and phase shifted to the various sub-fields of interest and averaged to a small field-of-view (typically 30-60\arcsec). The result is a small ($\sim$GB) dataset per sub-field which is easily manageable and parallelisable when calibrating and imaging. 

By combining multi-phase centre correlation with advanced calibration techniques such as in-beam phase referencing \citep{garrett2001agn,Garrett:2005fj,LencVLBI2008} and multi-source self-calibration \citep{Middelberg:2013hy,radcliffeMSSC2016}, wide-field VLBI surveys of milliarcsecond scale extragalactic radio sources to $\mu$Jy flux densities have become increasing feasible \citep[e.g.][]{,Middelberg:2011bx,Middelberg:2013hy,chi2013deep,Morgan2013M31,CaoVLBI2014,deller2014mjive,2015MNRAS.452...32R,Ruiz:2017ur}. 

We here present a new wide-field VLBI survey targeting the well studied Great Observatories Origin Deep survey North (GOODS-N) field using the European VLBI Network (EVN). The GOODS-N field covers 160 arcmin$^2$ with complementary deep multi-wavelength data including \textit{Chandra}, \textit{Spitzer}, \textit{Herschel}, UBVRIJHK photometry and spectroscopy. 

Previous wide-field VLBI observations targeted the Hubble Deep Field (HDF) and Flanking Fields (HFF) for which the GOODS-N field encompasses. \citet{garrett2001agn} used the EVN to target MERLIN sources within a 3.5\arcmin~ radius from the EVN pointing centre to r.m.s. sensitivities of 33\,$\rm\mu Jy\,beam^{-1}$. This resulted in the detection of 3 sources. These observations were substantially expanded upon by \citet{chi2013deep} who used Global VLBI to target the 92 VLA-MERLIN sources of \citet{muxlow2005high} within a $10\arcmin \times 10\arcmin$ field to r.m.s. sensitivities of 7.3\,$\rm\mu Jy\,beam^{-1}$. This resulted in 12 compact radio source detections (including the 3 detected by \citet{garrett2001agn}) thus beginning the characterisation of the faint compact radio population in GOODS-N. However, these surveys were invariably limited because computational limitations prevented imaging of the entire primary beam at that time. Our survey aims to substantially expand upon this sample, encompassing and surpassing the field-of-view and sensitivities of previous VLBI surveys in GOODS-N by targeting sources within a $30\arcmin\times30\arcmin$ area to $1\sigma$ central sensitivity of $\sim 2\mbox{-}3~{\rm \mu Jy\,beam^{-1}}$ with the completion of this survey.
 
In this paper, we present our initial catalogue of the 31 compact sources detected in the first data release to a $1\sigma$ sensitivity of $\sim 9~{\rm \mu Jy\,beam^{-1}}$ (corresponding to $\sim$ 17.5hr on source) along with derived radio properties of these objects using complementary 1-2\,GHz VLA data. In paper II, we compare our VLBI-selected population to other AGN detection diagnostics used in other wavebands. A future publication, paper III, will describe the final data release which will include an additional 48 hours of observations which comprise of the first wide-field VLBI observations using a combined eMERLIN-EVN array.

For this paper, we adopt a spatially-flat 6-parameter $\mathrm{\Lambda CDM}$ cosmology with $H_0 = 67.8\pm0.9~\mathrm{km\,s^{-1}\,Mpc}$, $\Omega_{m} = 0.308\pm0.012$ and $\Omega_{\Lambda} =0.692\pm0.012$ \citep{Planck2016}. We assume $S_\nu \propto \nu^{\alpha}$ throughout, where $S_\nu$ is the radio integrated flux density and $\alpha$ is the intrinsic source spectral index.

The paper is organised as follows. Section 2 outlines our observations, source selection strategy, calibration and source detection methodology. Section 3 details the primary beam correction method used for the EVN. Section 4 describes the VLBI catalogue accompanying this paper while a formatted version is presented in Tables 2 and 3.  Section 5 presents our results and associated discussion, including redshifts, astrometry, comparisons with other VLBI surveys and the radio properties of the VLBI-selected population. We conclude our findings in Section 6.

\section{Observations \& data reduction}\label{Sec: observations and data reduction}

\subsection{EVN observations}\label{SSec:Observations}
The EVN observed the \object{GOODS-N} field at 1.6 GHz for 24 hours in total on the 5\mbox{-}6th June 2014 (EVN code EG078B). The pointing centre used was the Hubble Deep Field-North (HDF-N), J2000 12:36:50.0 62:12:58.0. Ten telescopes, including the 100m Effelsberg and the 76m Lovell (Jodrell Bank) telescopes, were involved in the observation. In order to attain a uniform sensitivity profile, the Effelsberg and Lovell telescopes were nodded between 5 different pointing centres over the course of the observation while smaller telescopes remained pointed at the HDF-N centre. The observing strategy and participating telescopes are summarised in Table~\ref{table:telescopes}. 

These data were recorded at a bit rate of 1024 Mbits/s (8 $\times$ 16MHz bands) in both right and left hand circular polarizations. The fringe finders used were 3C345 and DA193. The observations were made using the standard phase referencing mode. Two phase calibrators were used; a strong, $\sim$0.4\,\mbox{Jy}, primary calibrator J1241+602 lying approximately $2^\circ$ from the target centre, and a weaker, 17\,\mbox{mJy}, secondary calibrator J1234+619 lying 23.5\arcmin~from the target centre. The primary calibrator was observed for 1.5 min on source every $\sim$$27$ minutes. To permit more accurate phase corrections, the secondary calibrator was observed more frequently (1.5 min on source every $\sim$$7.5$\,min).  In total, the on-source integration time on the GOODS-N field was approximately 17.5 hr.

\begin{table}
	\caption{EG078B observing strategy}             
	\label{table:telescopes}      
	\centering  
	\begin{tabular}{lll}
		\hline\hline 
		Telescopes & Country & Diameter (Derived) / m  \\
		\hline
		Ef & Germany & 100 (78) \\
		Wb & Netherlands & 25 \\
		On & Sweden & 25 \\
		Nt & Italy & 32 \\
		Tr & Poland & 32 \\
		Sv & Russia & 32 \\
		Bd & Russia & 32 \\
		Zc & Russia & 32 \\
		Sh & China & 25 (22.5)\\
		Jb1 & United Kingdom & 76 (67)\\
		\hline
		 \noalign{\smallskip}
	\end{tabular}  
	\begin{tabular}{llll}
		\hline\hline 
		Target Fields & R.A. (J2000) & Dec. (J2000) & Telescopes \\
		\hline
        HDF-N & 12:36:50.0 & +62:12:58.0 & All \\
        EFJB-P1 & 12:37:20.0 & +62:16:28.0 & Ef, Jb1 \\
        EFJB-P2 & 12:36:20.0 & +62:16:28.0	& Ef, Jb1 \\
        EFJB-P3 & 12:36:20.0 & +62:09:28.0 & Ef, Jb1 \\
        EFJB-P4 & 12:37:20.0 & +62:09:28.0 & Ef, Jb1 \\
        \hline
	\end{tabular}
	\tablefoot{{\it Upper panel:} Telescopes used in the EVN observations. Abbreviations, Ef: Effelsberg, Wb: WSRT (single dish), On: Onsala, Nt: Noto, Tr: Torun, Sv: Svetloe, Bd: Badary, Zc: Zelenchukskaya, Sh: Shanghai, Jb1: Jodrell Bank (Lovell Telescope). The fitted HPBW of telescopes with primary beam estimates are in brackets. \\{\it Lower panel:} Pointing centres used for the duration of the observation. Ef and Jb1 were nodded between all 5 pointing centres, whilst the rest of the array were pointed at the HDF-N pointing centre.}
\end{table}
\subsubsection{Source selection \& correlation}\label{SSec:Source selection and correlation}
These data were correlated using the SFXC correlator \citep{keimpema2015sfxc} at the Joint Institute for VLBI ERIC, Dwingeloo, the Netherlands. The correlation implemented the multiple simultaneous phase centre observing technique \citep[see][]{deller2007difx, deller2011difx, keimpema2015sfxc} to target 699 sub-fields/phase centres. Two source selection strategies were implemented and the criteria are as follows:

\begin{enumerate}[label=(\alph*)]
	\item A survey designed to completely cover the central 7.\!\arcmin5 radius area with VLBI coverage whilst restraining bandwidth and time smearing to $<$10\%. This is designed to complement the 1.5\,GHz eMERGE survey (Muxlow et al. in prep.) for use in integrated imaging. This comprises of 582 phase centres split into three main categories:
	
	\begin{enumerate}[label=\roman*.]
		\item 339 1.5\,GHz eMERLIN-VLA sources complete to \textasciitilde$11\,\mathrm{\mu Jy\,beam^{-1}}$ from the upcoming eMERGE survey (Muxlow et al. in prep.).
		\item 151 SCUBA-2 850\,$\rm \mu m$ sources (Smail priv. comm.).
		\item 92 other positions, denoted `legacy', which cover gaps in coverage across the field. 
	\end{enumerate}

	\item A targeted survey of 117 1.4\,GHz VLA radio-bright sources outside the central 7.\!\arcmin5 radius area with integrated VLA flux densities,  $S_{\mathrm{i,VLA}}$, $>0.1~\mathrm{mJy}$ \citep{morrison2010very}. These are split into:
	\begin{enumerate}[label=\roman*.]
		\item 91 sources with $0.1 < S_{\mathrm{i,VLA}} < 1$ mJy targeted to a radius of 12\arcmin~from the central pointing centre.
		\item 26 sources with $ S_{\mathrm{i,VLA}} > 1$ mJy targeted to the edge of the \citet{morrison2010very} VLA survey.We note that 4/26 of these sources are within the 15\arcmin and were accidental double entries in the correlation catalogue. These are kept in for clarity in-case these data are re-reduced in the future.
	\end{enumerate}
\end{enumerate}  

\begin{figure}[!tb]
	\centering
	\includegraphics[width=\hsize]{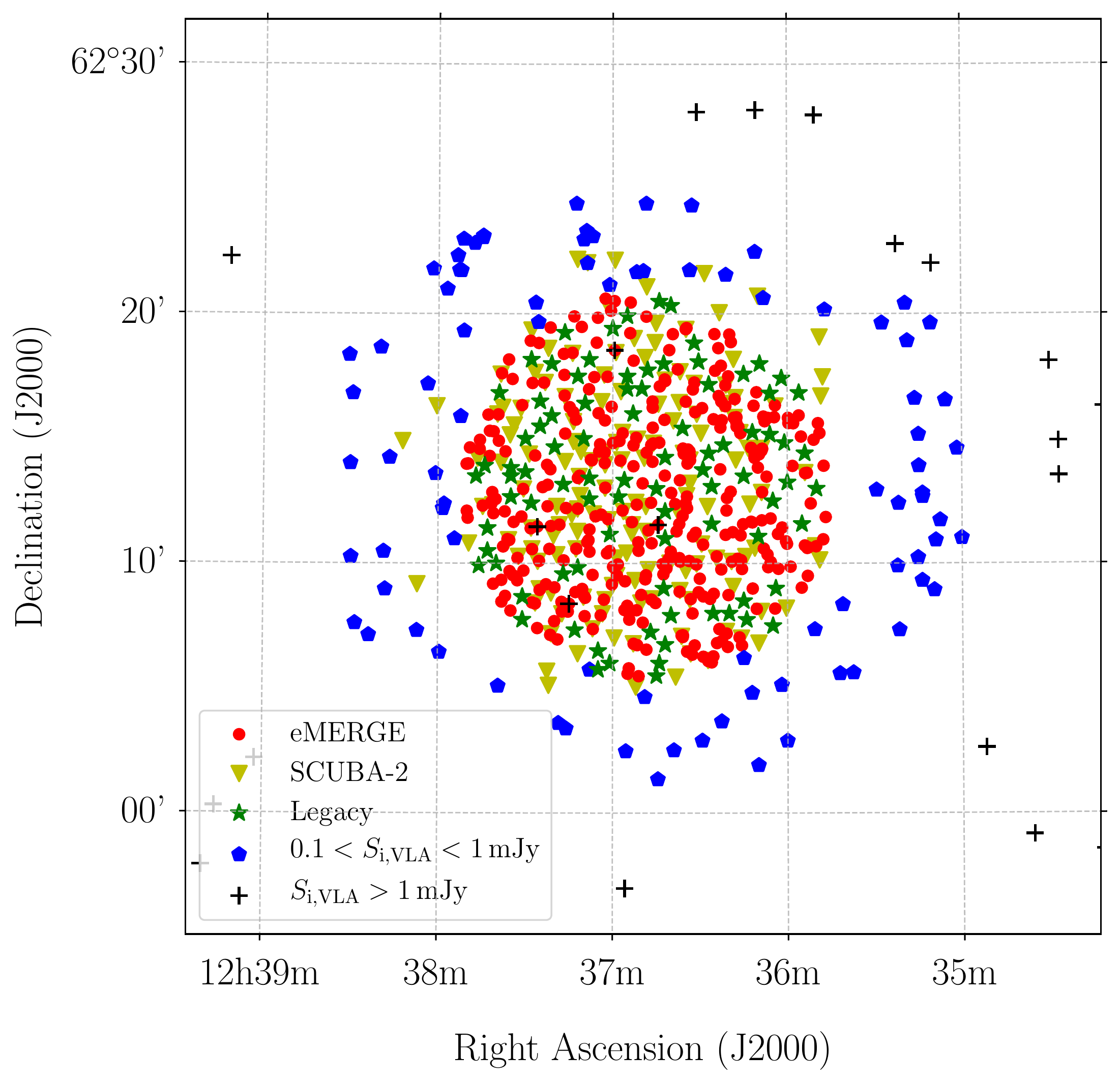}
	\caption{Sources /  sub-fields targeted by these observations. The central 7.\!\arcmin5 radius area complements the eMERGE survey and targets eMERLIN detected sources (red circles), SCUBA sub-mm sources (yellow inverted triangles) and legacy positions (green stars) which aim to fill in the gaps in coverage. The outer annulus targets only the brightest sources detected by the VLA in \citet{morrison2010very}. Those with integrated flux densities $0.1 < S_{\mathrm{i,VLA}} < 1$ mJy (blue heptagons) are targeted to a radius of 12\arcmin~and bright sources $S_{\mathrm{i,VLA}} > 1$ mJy (black crosses) are targeted to the edge of the \citet{morrison2010very} survey.
	}
	\label{Fig:Source positions}
\end{figure}

\noindent The source positions targeted are shown in Figure~\ref{Fig:Source positions}. With source positions determined, correlation proceeded as follows. Short sub-integrations of data were correlated at the required high spectral and temporal resolutions in order to reduce time and bandwidth smearing. In this observation, each sub-integration had a frequency resolution of 1.953\,kHz and a time resolution of 13.056\,ms in order to restrain time and bandwidth smearing to below 1\% on the longest baseline ($\sim 8400 $ km) at 5\arcmin\,from the pointing centre. At the end of each sub-integration, the visibilities were phase shifted to every desired source position to create a separate data set per position. Each data set was averaged to a temporal resolution of 3s and a frequency spacing of ${\rm 0.5\,MHz}$ (corresponding to a 10\% time and bandwidth smearing at 30\arcsec\,from the assigned source position) and then added to previous sub-integrations until the entire data set was correlated. As a result, 699 separate, narrow field-of-view (FoV) data sets were produced, one per source position. Attached to one data set, containing source J123462+621331, were the scans of the phase calibrators, J1241+602 and J1234+619, and the fringe finders, 3C345 and DA193 used for calibration. The phase referencing calibration and flagging tables derived for this data set can then be easily copied to the other data sets. Despite the total size being 3.79\,TB, calibration of this data set is easily parallelised and computationally less intensive than previous wide-field VLBI projects \citep[e.g.][]{chi2013deep}. For further clarification, we will refer to the target field as the GOODS-N field as a whole, while the sub-fields are the small FoV phase centres within the GOODS-N field whose coordinates were correlated upon.

 \begin{figure}[!tb]
	\centering
	\includegraphics[width=\hsize]{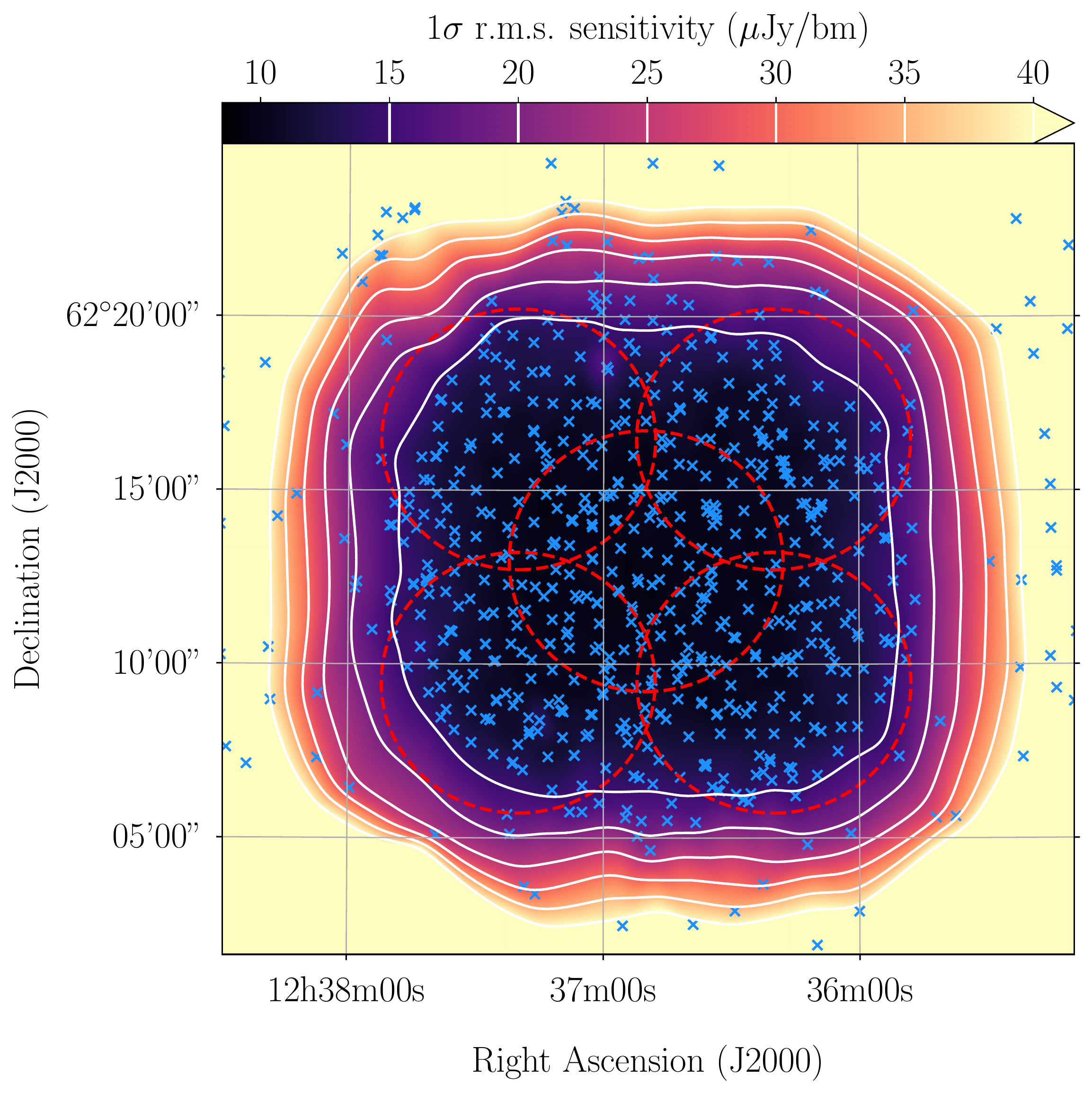}
	\caption{R.m.s. sensitivity for our 1.6~GHz EVN observations after primary beam correction. These data were optimally weighted for sensitivity (AIPS task \texttt{IMAGR: UVWTFN=`NA'}). The central r.m.s. is approximately \textasciitilde9~$\mu \mathrm{Jy\,beam^{-1}}$. The red dashed circles correspond to the HPBW of the Effelsberg telescope at 1.6\,GHz ($\sim7.\!\arcmin5$) at the pointing centres specified in Table~\ref{table:telescopes}, and coloured markers correspond to the sub-fields. Contours start at 15\,$\rm \mu Jy\,beam^{-1}$ in increments of 5\,$\rm \mu Jy\,beam^{-1}$ in order to illustrate the rapid sensitivity losses outside the primary beams of the large telescopes.}
	\label{Fig:rms_natural}
\end{figure}

\subsubsection{Data reduction}\label{SSec:Data reduction}

These data were reduced using the Astronomical image processing (AIPS) software developed by NRAO\footnote{www.aips.nrao.edu} \citep{Greisen:2003wo}, and its Python interface, Parseltongue \citep{kettenis2006parseltongue}. 

Before describing the data reduction, we note that there was an error found in the position of the secondary phase calibrator (J1234+619) when we tested phase referencing from the primary calibrator (J1241+602) to the secondary phase calibrator. This incorrect position originated from the \citet{chi2013deep} observations. The correct position was found to be J2000 12:34:11.7413(57) +61:58:32.478(07). Independent 5\,GHz e-MERLIN observations of J1234+619 using multiple phase calibrators verified that this new position is correct (see Appendix~\ref{appendix}). The model of J1234+619 derived from initial phase referencing tests were then used when fringe fitting in order to ensure the correct position is used.


With this issue established and solved, these data were calibrated as follows. Gains were calibrated using the system temperature, $T_{\mathrm{sys}}$, measurements from each antenna and the data were edited to remove any Radio Frequency Interference (RFI) using the AIPS tasks \texttt{SPFLG} and \texttt{CLIP}. Instrumental phase offsets between the spectral windows\footnote{We use the term `spectral windows' to describe the sub-bands in frequency. They are synonymous with the term IFs used in the AIPS data reduction package.} were then removed by solving for the phase and delays on a two minute integration of 3C345, using the task \texttt{FRING}. This allowed the spectral windows to be combined when the data is fringe fitted. We note that the dispersive delays were not corrected for, however we are confident that this is a minimal contribution as the fully calibrated data shows no phase deviations across the frequency band in excess of 10\mbox{-}15 degrees on all baselines. 

The group delays for the phase calibrators and fringe finders were calibrated using \texttt{FRING} (using a model of J1234+619 created when investigating the positional offsets), edited (using \texttt{SNEDT}) and smoothed (using \texttt{SNSMO}) to remove noisy and spurious solutions. With the delays calibrated, the phase and rates were then calibrated, edited and smoothed and applied to the data. We note that fringe fitting was conducted on both phase calibrators. This is because the Lovell Telescope (Jb) did not observe the primary phase calibrator due to a restriction on the number source changes per hour. With fringe fitting complete, the bandpass response was calibrated using AIPS task \texttt{BPASS}. 3C345 was used for bandpass calibration on all telescopes.

The primary phase calibrator (and furthest from the target), J1241+602, underwent three rounds of phase only self calibration (with solution intervals of five, one, and one minute(s) respectively) and one round of amplitude and phase self-calibration with a five minute solution interval. These solutions were applied to the closer, secondary phase calibrator, J1234+619. Three rounds of phase only self-calibration (with solution intervals of five, four, and two\,minutes respectively) were conducted on J1234+619 and these solutions were then applied to the sub-field containing J123462+621331. Spectral windows were not combined when self-calibration was performed on J1241+602, or during the first round of self-calibration on J1234+619. This would effectively correct any residual dispersive delay errors caused by a variable ionosphere by approximating the true phase correction (smoothly variable with frequency) with one that is a stepwise constant (one value per subband per solution interval).

The calibration solutions and flagging tables derived and applied to J123642+621331 were then applied to the other 698 sub-fields. All sub-fields were then imaged using AIPS task \texttt{IMAGR} using both natural (\texttt{UVWTFN=`NA'}) and uniform weighting schemes and these images were searched for emission. A detection threshold of $6\sigma$ was used to reduce the chance of false positives \citep[see][for more in depth discussion]{radcliffeMSSC2016}.

To reduce residual phase errors arising from atmospheric inhomogeneities between the phase calibrator and target field, we utilised the Multi-source Self-calibration (MSSC) technique developed by \citet{2013A&A...551A..97M} and \citet{radcliffeMSSC2016}. The nine brightest sources were used in MSSC. These sources were detected when imaged with both uniform and natural weighting schemes. If a source was detected in both images, it is highly suggestive that the source can be detected on all baselines. As a conservative precaution, we excluded sources outside the primary beam of the large telescopes (Effelsberg and Lovell) because we would expect considerable phase and gain errors to be induced by the attenuation of the primary beam. These errors would not be the same for each sub-field and will simply add noise into the solutions for MSSC when the individual sub-fields are combined. By performing this we ensure that the dominant error is from differing atmospheric paths between the phase calibrator and the target field. Three rounds of phase-only MSSC were performed using a solution interval of 2 min and the corrections derived were applied to all sub-fields. A primary beam correction scheme (outlined in Section~\ref{SSec:PBCOR}) was then applied to the central 7.\!\arcmin5 radius field and the sub-fields were searched for emission. The following section describes the method used to detect sources once MSSC and the primary beam correction was applied.

\begin{figure*}[!tb]
	\centering
	\includegraphics[width=\hsize]{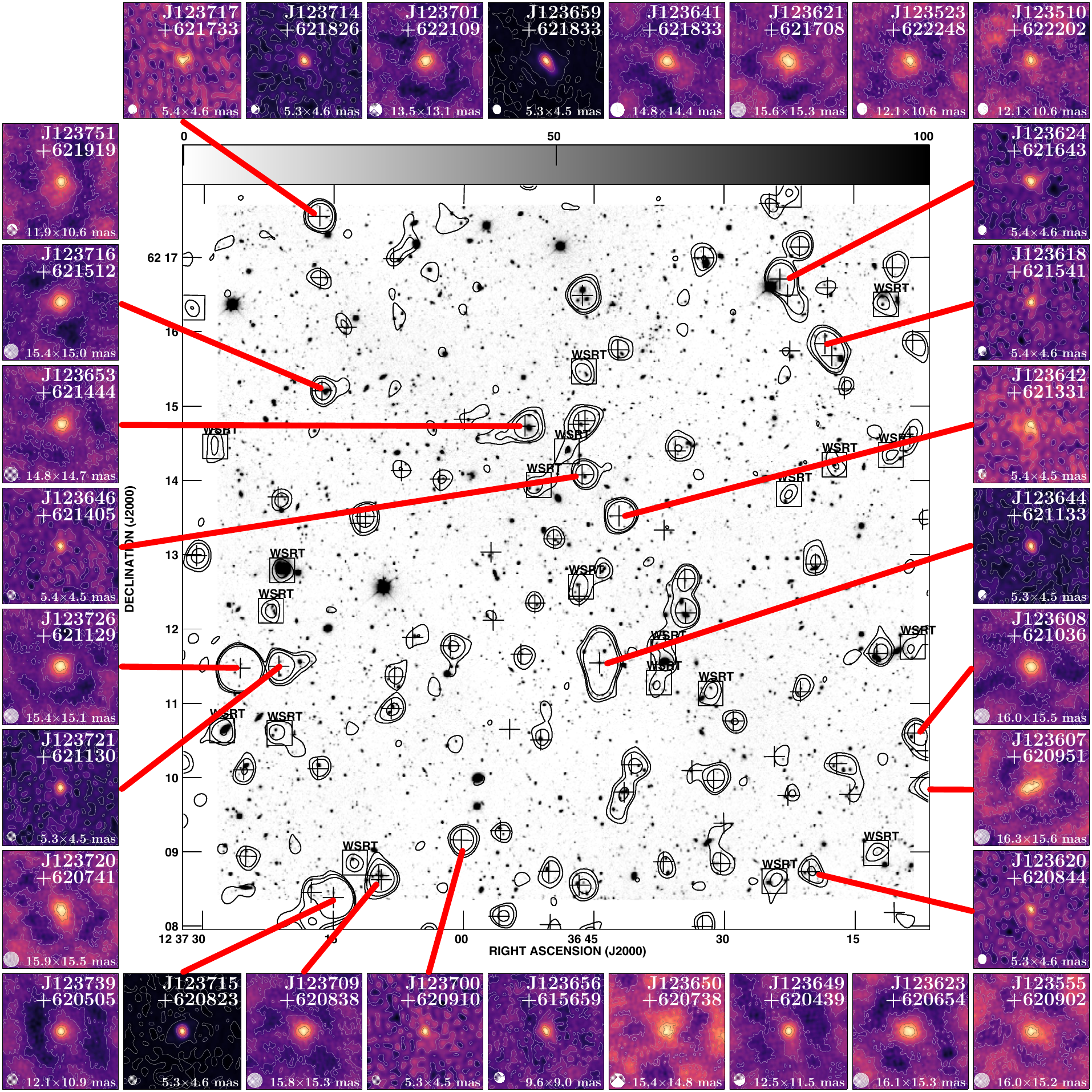}
	\caption{Composite image of 1.4\,GHz WSRT radio-KPNO optical overlay of the GOODS-N field, centred on the HDF-N \citep{garrett2000wsrt}, surrounded by postage stamp images of the 1.6\,GHz 31 VLBI detected sources presented in this paper. Those VLBI sources without adjoining red lines are located outside the WSRT central figure. The VLBI contours are $\pm1\sigma$ noise and then linearly spaced between $1\sigma$ noise and the peak pixel brightness. This image is an update on Figure 1 from \citet{chi2013deep}.}
	\label{Fig:All_sources}
\end{figure*}

\subsection{Source detection methodology}\label{SSec:source detection}

To determine accurate peak brightnesses and integrated flux densities, we tested multiple source detection algorithms namely AIPS task \texttt{SAD}, \texttt{BLOBCAT} \citep{hales2012blobcat} and \texttt{PYBDSF} \citep{Mohan2015:py}. It has been noted that Gaussian fitting routines, namely \texttt{SAD} and \texttt{PYBDSF}, were found to routinely over estimate integrated flux densities in the low signal-to-noise (S/N) regime where noise fluctuations across the extent of a source can induce sub-optimal fitting \citep[see][]{Middelberg:2013hy}. In the low S/N regime (S/N $\sim 6\mbox{-}10$), the measured integrated flux densities were on average $\sim16\%$ and $\sim22\%$ higher than the \texttt{BLOBCAT} measured values when using \texttt{SAD} and \texttt{PYBDSF} respectively. In the high S/N regime, the effects are less pronounced with both \texttt{SAD} and \texttt{PYBDSF} reporting fluxes only 4\mbox{-}5\% larger than \texttt{BLOBCAT}. 

We therefore used \texttt{BLOBCAT} to measure peak brightnesses and integrated flux densities of our sources. Initially, \texttt{BANE} \citep{Hancock2012:ag} was used to generate a r.m.s. map of each field which is in turn input to \texttt{BLOBCAT}. All parameters were set to default apart from the surface brightness error which was assumed to be $\sim$ 10\% which is caused by amplitude calibration errors (\texttt{-{}-pasbe=0.1}), the minimum S/N detection threshold (\texttt{-{}-dSNR = 6}) and, as our point spread function (psf or synthesised beam) is vastly oversampled, the peak brightness pixellation error was set to 1\% (\texttt{-{}-ppe=0.01}; see Appendix A of \citet{hales2012blobcat} for more information). The surface brightness error included an additional error factor which is proportional to the value of the primary beam correction. Note that \texttt{BLOBCAT} does not provide any size information therefore source sizes were measured using \texttt{PYBDSM}\footnote{As part of these observations we have developed a generalised wrapper for source detection in multi-phase centre VLBI observations that is publicly available. It can generate catalogues using \texttt{SAD}, \texttt{BLOBCAT} and \texttt{PYBDSF} (see \url{https://github.com/jradcliffe5/General_VLBI_cataloger})}. 

For the central 7.\!\arcmin5 radius field, each sub-field was imaged using natural weighting only (to optimise sensitivity) and then the method outlined above was used to search for detections. Detections were then imaged with uniform weighting (whose r.m.s. is $~1.6\times$ the naturally weighted r.m.s.) in order to obtain the highest resolution image possible.
	
To optimise the number of detections, we used a different strategy to image sub-fields outside the central  7.\!\arcmin5 radius field which are beyond the half-power beam widths (HPBWs) of the large telescopes (Effelsberg and Lovell). Two images were made for each sub-field. For the first image, the large telescopes on all pointings apart from the closest pointing centre to the sub-field were flagged. This was performed because while the large telescopes retain significant sensitivity well beyond the HPBW of their primary beam, the retention of the more distant pointing centres would induce significant amplitude errors which would outweigh any sensitivity gains. This method produced an additional 6 detections, the majority of which (4/6) are within a 12\arcsec~radius of the pointing centre. For the second image, all of the large telescopes are flagged, so that sources up to the HPBW of the smaller 32m and 25m telescopes could be detected without being affected by amplitude errors from the large telescopes still present in these data. This method produced just one additional detection and none of the 6 sources detected with the larger telescopes included were detected with this method. This is likely due to the significant sensitivity reduction when the large telescopes are removed. Additionally, we note that the primary beam models are poorly constrained outside the HPBW, therefore these sub-fields do not have primary beam correction applied.

Once detections were identified, each sub-field was re-imaged with both uniform weighting ($\sim5.3\times4.5$ mas) and natural weighting ($\sim 16\times16$ mas) schemes and re-catalogued resulting in a total of 31 detections (24 from the central field and 7 from the targeted survey beyond the 7.\!\arcmin5 radius field). These detections are shown in Figure~\ref{Fig:All_sources} which is an update on Fig. 1 from \citet{chi2013deep}.  We note that \citet{radcliffeMSSC2016} only stated an initial 20 sources, however this study invariably missed detections because the majority of the annulus sub-fields were not included. The derived peak brightnesses, flux densities and positions of our objects are described in Table ~\ref{Table:Source_catalog}.

\subsection{VLA observations}\label{SSec:VLA Observations}
In addition to the EVN observations, archival L-band Karl. G. Jansky Very Large Array (VLA) A-array data (P.I. F. Owen) were reduced to provide a zero\mbox{-}spacing flux density for our VLBI sources and forms part of the eMERGE survey data (Muxlow et al. in prep.). This section briefly describes the data reduction process. The VLA observed the GOODS-N field between the 7th August-11th September 2011 for a total of 38 hours, in the A-array configuration. These data were flagged using the \texttt{AOFlagger} software \citep{Offringa:2012eo} and calibrated using the VLA CASA calibration pipeline\footnote{\url{https://casa.nrao.edu}} (packaged with CASA version 4.3.1). These data were transferred to AIPS and two sources (J123452+620236 and J123538+621932) were peeled. 

Due to the large fractional bandwidth ($\sim$68\%) and large data size, postage stamps centred on each VLBI detection were produced using the multi-term multi frequency algorithm within CASA task \texttt{tclean} \citep{Rau2011:mfs}. These images were primary beam corrected using the CASA routine \texttt{widebandpbcor}, which also corrects for the induced spectral index caused by the varying primary beam attenuation across the bandwidth. The resulting images have a r.m.s. of $\sim 2\mbox{-}5~\mathrm{\mu Jy\,beam^{-1}}$ with a restoring beam of $1.54\arcsec\times1.34\arcsec$. Flux densities were extracted using \texttt{BLOBCAT} \citep{hales2012blobcat} and we conservatively assign a standard 10\% surface brightness error originating from calibration. 

\section{EVN primary beam correction}\label{SSec:PBCOR}

For these observations, we used and developed one of first primary beam models European VLBI Network (Keimpema et al. in prep.). We followed a similar prescription to primary beam modelling as described in \citet{Strom2004:PB} and \citet{CaoVLBI2014}. Due to the lack of accurate primary beam models for many EVN telescopes, the primary beam power response of each telescope can be approximated by using a normalised, symmetric, 2D Gaussian of the form,

\begin{equation}
P(\theta, \phi) \approx \exp\left(-{\frac{(\theta- \theta_0)^2 + (\phi - \phi_0)^{2}}{2\sigma^2}}\right),
\end{equation}

\noindent where $P(\theta,\phi)$ is the relative primary beam power response. $\theta$ and $\phi$ are the respective azimuthal and polar angular distances from the antennas' pointing centres. The azimuthal and polar coordinates of the telescope's pointing centers are defined by $\theta_0$ and $\phi_0$ respectively. The standard deviation, $\sigma$, can be related to the FWHM of the primary beam, $\theta_{1/2}$, through the expression,

\begin{equation}
\sigma^2 = \frac{\theta_{1/2}^2}{8\ln{2}},
\end{equation} 

\noindent where the FWHM of the primary beam is defined as,
\begin{equation}\label{eqn. beam}
\theta_{1/2} = K\frac{\lambda_{\mathrm{c}}}{D}. 
\end{equation}

\noindent Here, $\lambda_{\mathrm{c}}$, is the observing wavelength, and $D$ is the aperture diameter (Keimpema 2015, unpublished). A small correction factor, $K = 1.05$, was used to take into account any aperture blockages (Wrigley et al., in prep.). For some telescopes (namely Effelsberg, Jodrell Bank and Shenzhen in these observations) accurate beam models are available and therefore replace $D/K$ in Eqn.\,\ref{eqn. beam} with the fitted HPBW of these telescopes. We note that there are some uncertainties associated with the derived HPBW of Jodrell Bank because this is derived from a modelled aperture distribution rather than more accurate methods such as holographic scans \citep{Wrigley:thesis}. The fitted aperture diameters are summarised in Table~\ref{table:telescopes}.

$P(\theta, \phi)^{-1/2}$, that is the inverse primary beam voltage response, was calculated for each telescope at every time integration step, each spectral window and each sub-field. These were recorded into an AIPS SN table (one per sub-field) which were then be applied to the corresponding $uv$ data set using AIPS task \texttt{CLCAL}. The application of this calibration table multiplies the visibility amplitudes of each baseline by a correction factor which is the inverse product of the primary beam voltage responses, ($P_i(\theta, \phi)^{-1/2} \times P_j(\theta, \phi)^{-1/2}$) of the two telescopes $i,j$ that form the baseline. The application of this table also adjusts the weights by the inverse of this correction factor. The simultaneous adjustment of weights means this primary beam correction will also correctly weight multiple pointings, thus permitting mosaicking using the EVN array.

We note that the use of a Gaussian model for our primary beam model implies heavy tapering of EVN telescopes, however these telescopes were designed for single dish observations so we would expect a large amount of tapering. In addition, we tested multiple models (1D Gaussian, 2D Gaussian, Airy disk, and a polynomial) when fitting to these beam models and found that differences between the models are only significant towards the primary beam null and these models are indistinguishable within the HPBW.

However, because of the lack of available primary beam models for many EVN telescopes (most notably the Lovell telescope for this observation), there are considerable uncertainties on the derived peak brightnesses and integrated fluxes. Incorporating this model to include higher order corrections, such as beam squint and more physically motivated models would be simple. Nevertheless, if we are to significantly improve EVN primary beam correction, beam measurements for all individual EVN stations are required, because uncertainties towards the edge of the primary beam are dominated by the lack of information on the primary beams' HPBWs.

As a result of this, sub-fields which lie outside the central 7.\!\arcmin5 radius area were not primary beam corrected because, the Effelsberg and Lovell telescope corrections are beyond the FWHM of the Gaussian models derived, and errors due to incorrect beam models will rapidly increase. Figure~\ref{Fig:rms_natural} shows the r.m.s. sensitivity of our observations after primary beam correction using natural weighting. The central r.m.s. is approximately 9 $\mathrm{\mu Jy\,beam^{-1}}$.

We note that this primary beam correction is constantly under-development and will be updated with the latest EVN beam models. The code is publicly available and can be found at \url{https://github.com/jradcliffe5/EVN_pbcor}.

\section{Catalogue description}\label{Sec:Catalog}
\begin{figure}[!tb]
	\centering
	\includegraphics[width=\hsize,clip]{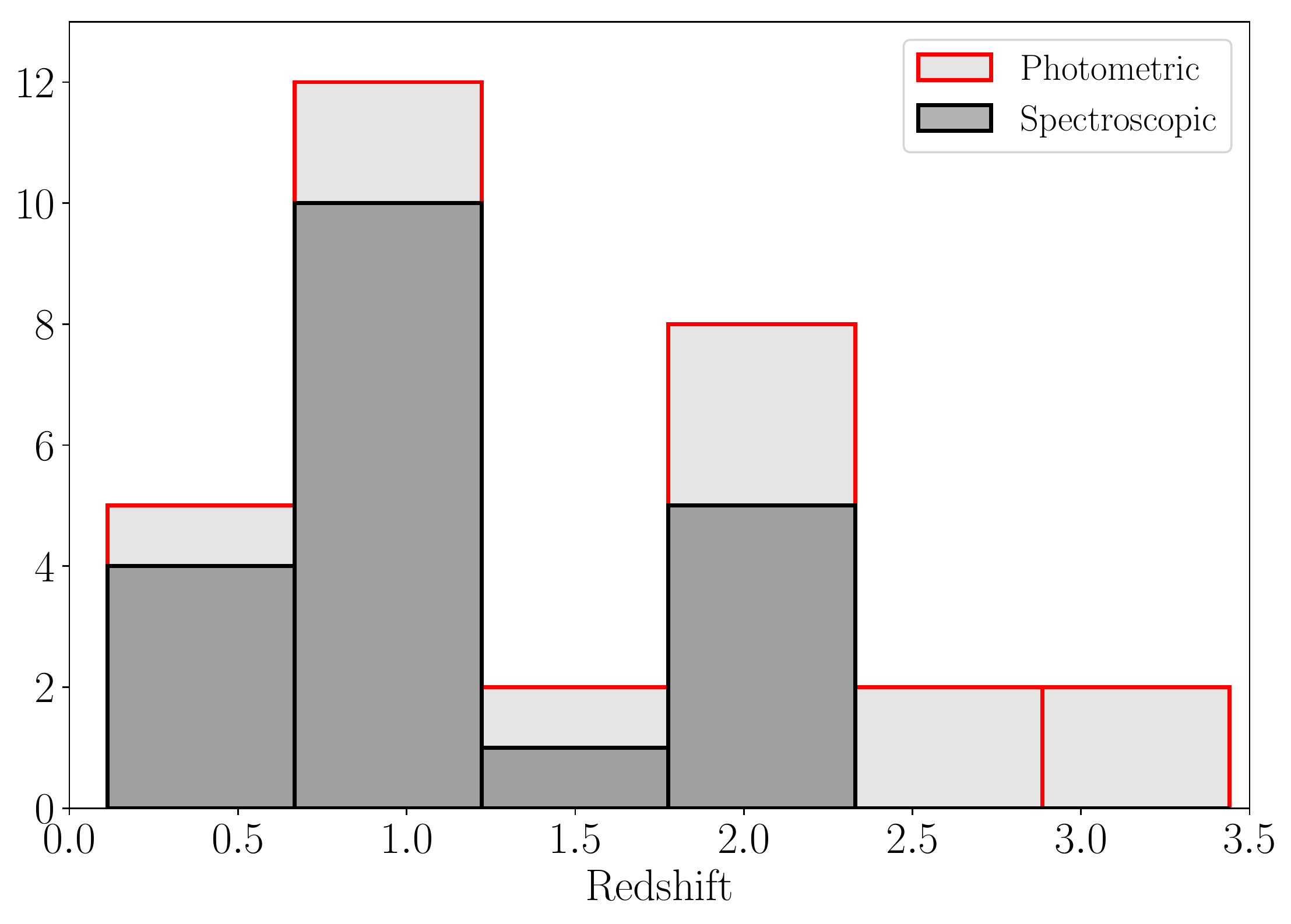}
	\caption{Redshift distribution for our detected VLBI sources. There are 20 spectroscopic redshifts (dark grey) and 11 photometric redshifts (light grey) for these objects. The over-density of sources at $z\sim2$ are briefly discussed in Section~\ref{SSec:redshifts}. Optimal bin widths were calculated using the prescription in \citet{Knuth2006:his}.}
	\label{Fig:Redshifts}
\end{figure}

In this section we describe the VLBI catalogue of 31 compact radio sources which accompanies this paper (see Tables~\ref{Table:Source_catalog} and \ref{Table:Derived_properties}). The designated column numbers correspond to the associated machine-readable version of the catalogue. The column descriptors are as follows:

\begin{enumerate}[leftmargin=2\parindent]
	\item[Col. 1:] \textit{Source ID}. Radio name adopted in this paper which is of the form Jhhmmss+ddmmss based upon the J2000 Right Ascension (in hours) and Declination (in degrees). Note that some source identifiers are slightly different to that of \citet{morrison2010very} because of the improved astrometric accuracy.
	
	\item[2-4:] \textit{z}.  The redshifts for the 31 VLBI detected sources can be found in Column 2. The 68\% lower and upper confidence intervals are in Columns 3-4. A description of how these redshifts were compiled can be found in Section~\ref{SSec:redshifts}.
	
	\item[5,6:] \textit{z type/ref}. Column 5 describes how these redshifts were determined. Spectroscopic redshifts are denoted with `S' in the $z$ type column while photometric redshifts are denoted with `P'. Column 6 contains the reference for which the redshift was acquired
	
	\item[7:] \textit{R.A. (J2000)} Right Ascension (J2000) in hours with the format hh:mm:ss.ssss. 
	
	\item[8:] \textit{Dec. (J2000)} Declination (J2000) in degrees with the format dd:mm:ss.sss.
	
	\item[9-10:] \textit{VLBI $P$}. Peak brightness (Column 9) and associated error in $\mu\mathrm{Jy\,beam^{-1}}$ (Column 10). This is determined using the flood-filling algorithm \texttt{BLOBCAT} \citep{hales2012blobcat}. Errors derive from source fitting, calibration, and primary beam correction. Peak brightnesses of those sources without primary beam correction are merely lower limits (denoted by -99.0 in the error column in the machine-readable table).
	
	\item[11-12:] \textit{VLBI $I$}. Integrated flux density (Column 11) and associated error in $\mu\mathrm{Jy}$ (Column 12). This is determined when deriving the peak brightness. Errors originate from source fitting, calibration, and primary beam correction. Integrated flux densities of those sources without primary beam correction are merely lower limits (denoted by -99.0 in the error column in the machine-readable table).
	
	\item[13:] \textit{S/N}. Signal-to-noise ratio.
	
	\item[14-16:] \textit{Beam}. Major axes (Column 14), minor axes (Column 15) and position angle (Column 16) of the CLEAN restoring beam in milliarcseconds. The two restoring beams of $\sim 16 \times 16$ mas and $\sim 5.3 \times 4.5$ mas correspond to natural weighting and uniform weighting schemes respectively. The VLBI peak brightnesses and integrated flux densities presented in Columns 9-10 and 11-12 respectively have been derived using the beam sizes specified in these columns.
	
	\item[17-18:] \textit{VLA $P$}. Peak brightness (Column 17) and associated errors (Column 18) of the 1.5\,GHz VLA  observations described in Section~\ref{SSec:VLA Observations}. Errors are determined from calibration ($\sim 10\%$) and source fitting.
	
	\item[19-20:] \textit{VLA $I$}. Integrated flux density (Column 19) and associated errors (Column 20) of the 1.5\,GHz VLA observations. Errors originate from calibration ($\sim 10\%$) and source fitting.
	
	\item[21:] \textit{$\alpha$}. Spectral index between the 5.5\,GHz VLA observations of \citet{Guidetti:2017wt} and 1.5\,GHz VLA observations presented in this paper. 
	
	\item[22-23:] \textit{$L_{\mathrm{1.5GHz}}$}. Monochromatic rest-frame radio luminosity in $\rm W\,Hz^{-1}$ (Column 22) and its associated error (Column 23). Values were derived using the 1.5\,GHz VLA integrated flux densities and k-corrected using the spectral index. Associated errors originate primarily from those sources with photometric redshift errors. The median spectral index of $-0.56$ were used to determine the radio luminosities of sources without spectral index information. 
	
	\item[24-26:] \textit{$T_b$}. Brightness temperatures, in $\rm K$, calculated using Eqn.~\ref{equation:brightness temperature} (Column 24). Column 25 is a flag to denote lower limits (set as 1 in machine readable version to denote lower limits) and Column 26 corresponds to whether the brightness temperature was calculated using uniform or natural weighting. This is denoted as U or N respectively in the machine-readable table and in Table~\ref{Table:Derived_properties} brightness temperatures derived using natural weighting are italicised. See Section~\ref{SSSec:Brightness temperatures} for further details.
	
	\item[27-30:] \textit{Angular sizes}. Columns 27 and 29 describe the major and minor axes of the deconvolved projected angular size of the VLBI source in milliarcseconds. Columns 28 and 30 contain flags to denote upper limits (hence unresolved source sizes) for the major and minor axes respectively (set as 1 in machine readable version to denote upper limits). Sizes were fitted using \texttt{PYBDSF} \citep{Mohan2015:py} and see Section~\ref{SSSec:Brightness temperatures} for further details.
	
	\item[31-34:] \textit{Linear size}. Columns 31 and 33 describe the  major and minor axes of the deconvolved projected linear size of the VLBI source in parsecs and Columns 32 and 34 contain flags to denote upper limits for the major and minor axes respectively (set as 1 in machine readable version to denote upper limits).
	
\end{enumerate}

\section{Results and discussion}

\subsection{Redshifts}\label{SSec:redshifts}
The VLBI positions were matched to the many spectroscopic and photometric catalogues to within a radius of one arc-second from the VLBI position. In order to prevent mis-identifications potential matches were visually compared to the HST optical/near-IR images of \citet{Skelton_HST3D_2014} to ensure the correct redshift is assigned. Redshift information was found for all 31 objects. This comprises of 20 spectroscopic redshifts \citep[][L. Cowie priv. comm.]{Cowie2004z,Smail2004:sz,Chapman:2005wh,Barger_specz_2008,Murphy:2017ja} and  and 11 photometric redshifts \citep{Skelton_HST3D_2014,Yang_photz_2014,Cowie2017z}. The redshift distribution is shown in Figure~\ref{Fig:Redshifts}. The median redshift is 1.146 and ranges from 0.11 to 3.44.

The redshift distribution shows an abundance of sources around $\sim 2$ which could be an imprint from the previously identified $z=1.99$ proto-cluster in GOODS-N \citep{Blain2004:smg,Chapman2009:sm}. This proto-cluster has approximate redshift bounds of $1.982 < z < 2.010$, and two VLBI sources (J123618+625541 and J123621+621708) are associated this proto-cluster \citep{Casey2016:sm}. It is expected that this structure is extended spatially beyond the limited field-of-view of the GOODS-N survey as the deep spectra does not extend much outside of the HST coverage. It has been suggested that J123642+621331 ($z=2.018$) could also member of this proto-cluster \citep{Murphy:2017ja}, and it is also possible that J123721+621130 ($z=2.02$) could be associated, however an accurate spectroscopic redshift would need to be acquired. This raises possibility that, with deep VLBI surveys and improved number densities of sources, over-densities of VLBI-detected AGN could act as a tracer of proto-clusters for which there is evidence of enhanced AGN activity in multiple wavebands including the radio \citep[e.g.][]{,Wylezalek2013:agn,Krishnan2017:pc}.

\subsection{Astrometry}
\defcitealias{muxlow2005high}{M05}
In order to check the astrometry of these observations, we compared these VLBI observations to the MERLIN-VLA observations from \citet{muxlow2005high}, hereafter \citetalias{muxlow2005high}), and the reprocessed wide-field images of the same data by \citet{Wrigley:thesis}. The positions were not compared to the global VLBI observations of \citet{chi2013deep} due to the known positional uncertainty on the phase calibrator used in those observations (see Appendix A), plus there is a larger number of concordant sources between these EVN observations and \citetalias{muxlow2005high}. Due to computational limitations at the time, \citetalias{muxlow2005high} targeted 92 radio sources within a $10\arcmin\times10\arcmin$~square field with VLA-only flux densities $> 40~\mathrm{\mu Jy}$ in the HDF-N field to a central r.m.s. of $3.3 \mathrm{\mu Jy\,beam^{-1}}$. 

These data were re-processed by \citep{Wrigley:thesis} and a primary beam corrected $18\arcmin\times16\arcmin$ image was used for the subsequent analyses. We used \texttt{BLOBCAT} (with a detection threshold of 6$\sigma$) to generate a catalogue of 155 MERLIN-VLA detected sources (with a restoring beam of 0.2\arcsec). These were then cross matched with the EVN positions to within 1 arcsecond. A total of 25/31 sources were matched to an MERLIN-VLA detected source with the remaining unmatched sources beyond the sky area considered by \citet{Wrigley:thesis}. We estimate a conservative 5 mas error for the astrometry of these new observations, due to uncertainties on the position of the secondary phase calibrator J1234+619, and a 10 mas error on the VLA-MERLIN data arising from calibration and source fitting errors. 

\begin{figure}[!tb]
	\centering
	\includegraphics[width=\hsize]{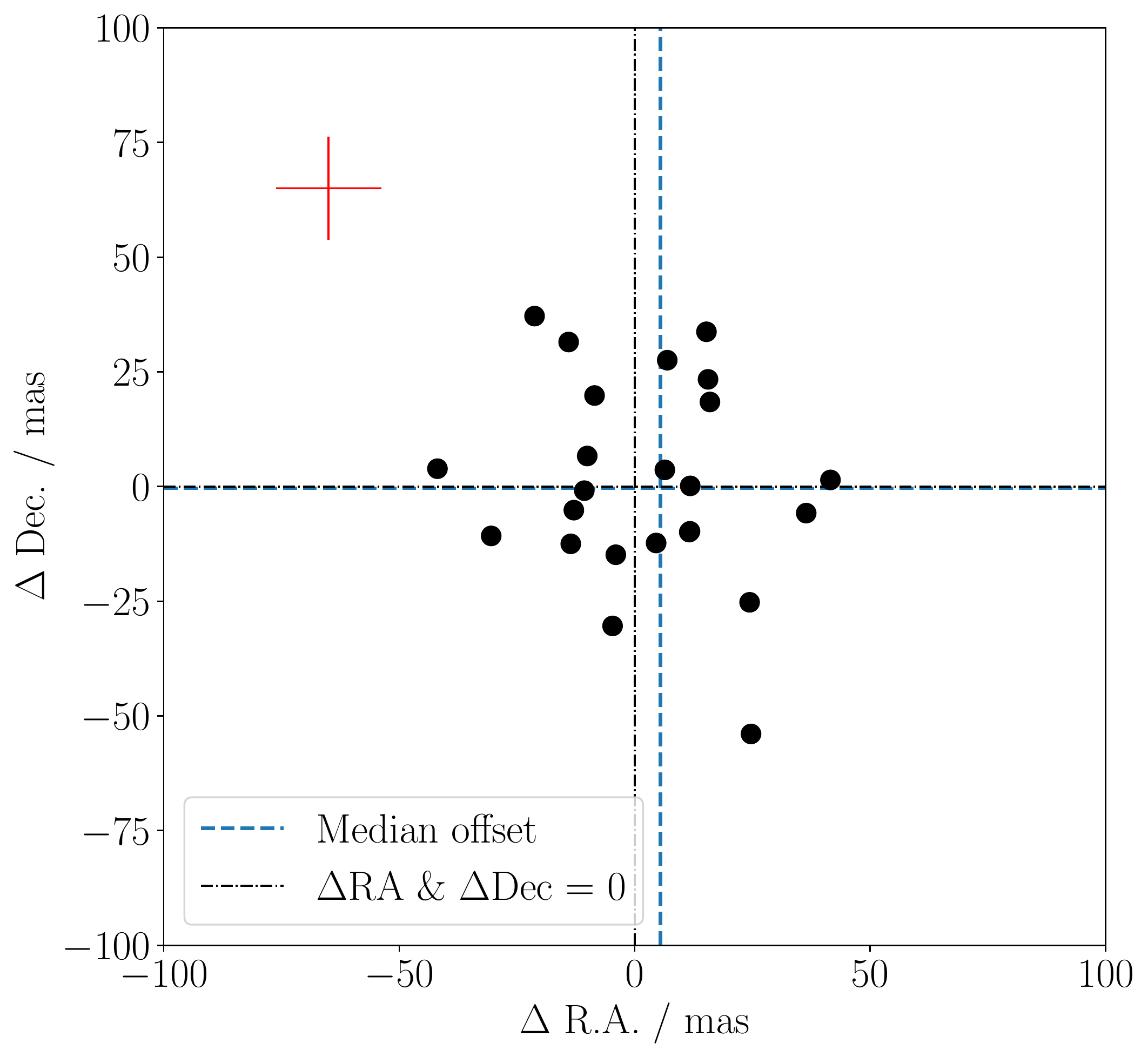}
	\caption{Relative astrometric precision between \citet{muxlow2005high} and these observations. The blue dashed line is the mean RA and Dec shifts corresponding to 5.5  and -0.4 milliarcseconds respectively. The black dot-dashed line corresponds to $\Delta$RA=0 and $\Delta$Dec=0. The red cross indicates the typical error per data point (which does not include core-jet blending uncertainties).}
	\label{Fig:astrometry}
\end{figure}

As Figure~\ref{Fig:astrometry} shows, there is a small systematic offset of 5.5 mas in RA and 0.4 mas in Dec. Note that all sources apart from wide-angle tail FR-I source, J123726+621129, are within 60 mas of the MERLIN-VLA positions. This source was excluded from the derivation of the systematic astrometric offset and Figure~\ref{Fig:astrometry}. The systematic offsets can originate from core-jet blending of the radio emission, standard errors associated with source fitting and calibration, and also errors originating from repeated use of the AIPS task UVFIX when peeling bright sources from the VLA-MERLIN data \citep[e.g. see][]{MorganVLBI2011}. The astrometric scatter is expected to be dominated by blending in the MERLIN-VLA data as substructure such as AGN jets will blend with AGN core emission thus causing an offset in position of peak brightness. These errors should have orientations which are randomly distributed, hence the median 5\,mas astrometry offset indicates that we are in fairly good agreement with the MERLIN-VLA positions.

\subsection{Comparison to other VLBI surveys}

As previously stated, the GOODS-N field has been targeted by two previous wide-field VLBI surveys by \citet{garrett2001agn} and \citet{chi2013deep}. It is worth noting that surveys had restricted field-of-views which are encompassed by the field-of-view of these new EVN observations, therefore we would expect that we should be able to detect all previous identified sources. This survey recovers 11/12 of the \citet{chi2013deep} detections and all three of the \citet{garrett2001agn} detections. The missing source, J123642+621545, illustrates significant radio variability and has an e-MERLIN integrated flux density of only 60 $\mu$Jy during the period of these observations, whereas \citet{chi2013deep} detects an integrated flux density of 343 $\mu$Jy.


We compared our results with other wide-field VLBI surveys to ensure that our observations and detection fractions are consistent. In order to calculate the detection fraction, we used the VLA-A array observations provided by \citet{morrison2010very} and cross-matched these with to our VLBI survey. For this analysis, we only consider the central area where we have contiguous imaging and where our primary beam model is most reliable. Within this region, our EVN data were used to image the locations of the 314 VLA sources.  Of these sources, 94 were theoretically detectable assuming that they are unresolved on VLBI angular scales with a flux density greater than our 6$\sigma$ VLBI detection threshold (based upon the VLBI r.m.s. noise distribution, as shown in Fig.~\ref{Fig:rms_natural}).  Of this sample 24 sources were detected with VLBI, thus giving a detection threshold of $25.5$$\substack{+5 \\ -4}$\% (24/94). Errors were determined using the Bayesian binomial estimator of \citet{Cameron2011:st}. This is consistent with previous wide-field VLBI surveys. For example, the Chandra Deep Field-South \citep[$55~\mu$Jy\,beam$^{-1}$ r.m.s.,][]{Middelberg:2011bx} has a detection fraction of $20\substack{+5 \\ -4}\%$, the Lockman Hole/XMM \citep[$24~\mu$Jy\,beam$^{-1}$ r.m.s.,][]{2013A&A...551A..97M} detects $30\pm3\%$ and the COSMOS survey \citep[$10~\mu$Jy\,beam$^{-1}$ r.m.s.,][]{Ruiz:2017ur} detects $20\pm1\%$. The mJIVE survey has a detection fraction of $20\pm1\%$ to a an r.m.s. of $60\mu$Jy$\,$beam$^{-1}$ \citep{deller2014mjive}.
	
Note that there are some caveats because our VLBI sample is surface brightness limited due to the resolution and flux sensitivity of our EVN observations. For the fainter sources in our targeted sample (e.g. $S_{\rm VLBI}$ $\lesssim$100 $\mu$Jy), we can only detect those VLA sources with relatively large VLBI-VLA flux density ratios. For example, as the median flux density ratio is \textasciitilde0.6 for this sample (see Section~\ref{SSec:vlbi_vla}), most VLA sources with integrated flux densities $\lesssim$ 100 $\mu$Jy will go undetected. This motivates deeper VLBI observations in the future as the improved r.m.s. noise levels will recover these sources.
	

\subsection{Radio properties of the VLBI selected population}\label{Sec:Radio Properties of the VLBI Selected Population}
\subsubsection{VLBI-VLA flux densities}\label{SSec:vlbi_vla}
We compared the 1.5\,GHz VLA flux densities to our 1.6\,GHz EVN observations in order to investigate the origin of the radio emission. To do this we can use the VLBI to VLA flux density ratio ($R$) to establish whether a source is dominated by milliarcsecond-scale emission from AGN cores or arc-second scale emission from AGN jets/lobes or star-formation related processes. Note that for this comparison, we only use the VLBI sources which have been primary beam corrected (24/31). As both the VLBI and VLA observations have complete 24hr $uv$ coverage, we use the integrated flux density measurements to define the VLA-VLBI flux density ratio ($R \equiv S_{\rm i,VLBI}/{S_{\rm i,VLA}}$).

We find that the 18/24 (66.7\%) VLBI sources have over 50\% of their radio emission originating from a milli-arcsecond scale component. 2/24 (8\%) has $R>1$ which is most likely due to AGN variability because the observation times of VLA (2011) and VLBI (2014) data vary by a few years. Excluding the two known variable sources with $R>1$, we find that the median VLBI-VLA ratio of our observations is 0.625. This is largely in agreement with the COSMOS VLBI survey which find a median VLBI-VLA ratio of 0.6 \citep{Ruiz:2017ur}. We note that, at low flux densities, our VLBI observations are expected to preferentially detect core-dominated systems, with the majority of VLA arcsecond-scale emission confined to a high brightness temperature core which is detectable by VLBI observations. This is consistent as only a small fraction ($\sim10\%$) of our VLBI-detected sources exhibit large-scale radio jets or lobes, while the remaining sources are compact. 


An evolution towards more core-dominated systems as we approach $\mathrm{\mu Jy}$ flux densities has been hinted at in other VLBI surveys, most notably the mJIVE-20 and COSMOS VLBI surveys \citep[][]{deller2014mjive,Ruiz:2017ur}. There is some evidence that suggests that this evolution may be true. It has been shown that a population of radio sources with core fraction of $\sim$0.3 below a 1.4 GHz luminosity of $10^{25}~\mathrm{W\,Hz^{-1}}$ are required by empirical simulations in order to accurately extrapolate the established populations from low-frequency ($< 5$\,GHz) surveys to the $>$ 10\,GHz source populations \citep{Whittam:2017sk}. This could be equivalent to the postulated population of `FR0' sources in the local universe which have core dominated, compact radio emission extending to at most just 3\,kpc \citep{Baldi:2015jt}. These wide-field VLBI surveys could be beginning to detect the high-$z$ analogues to this population of radio sources.

\subsubsection{Luminosities}\label{SSec:Luminosities}
\begin{figure}[!tb]
	\centering
	\includegraphics[width=\hsize]{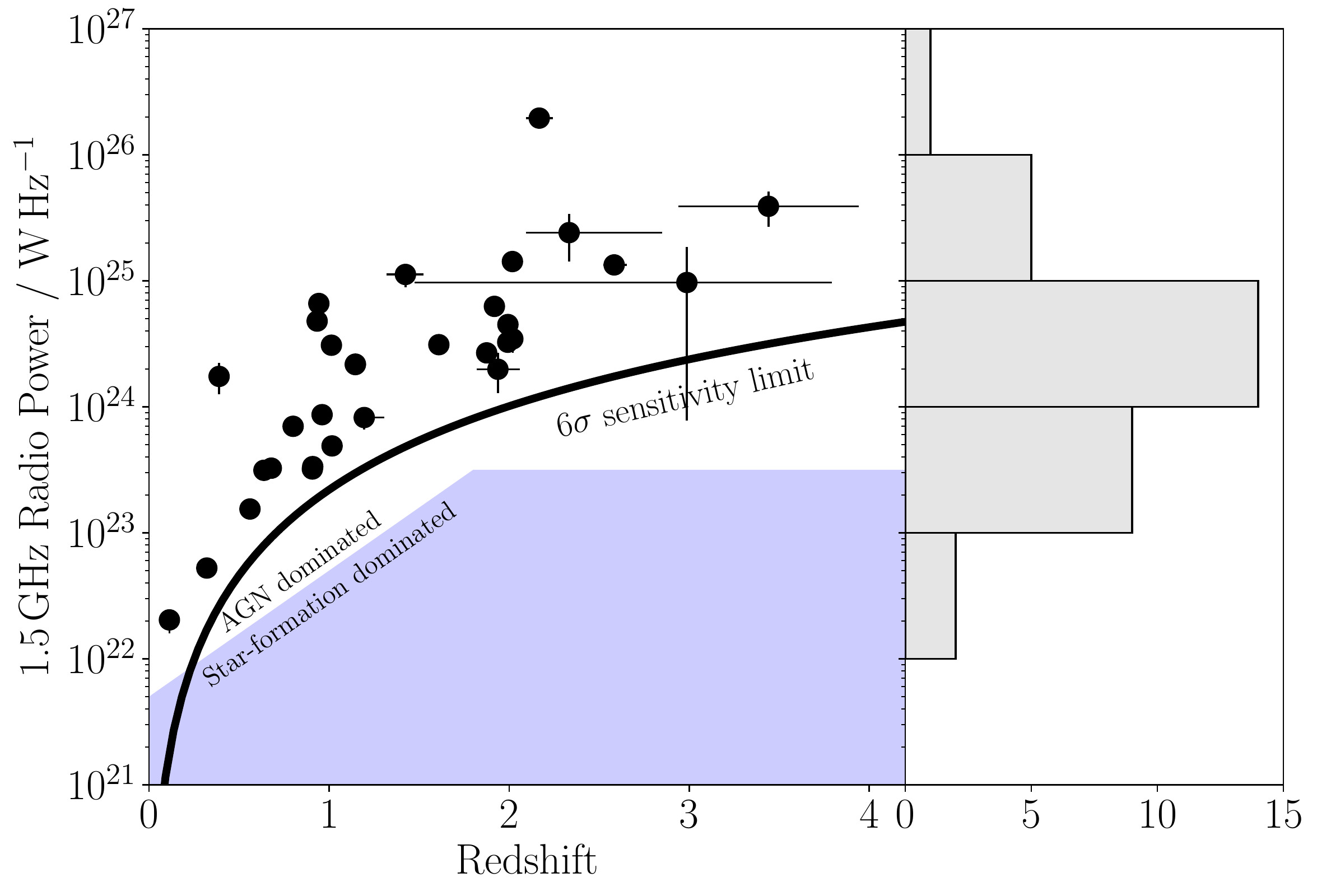}
	\caption{Radio power vs. redshift for our VLBI sources. 1$\sigma$ uncertainties on radio power and redshifts are plotted. The bold black curve represents the theoretical radio power that these VLBI observations are sensitive to (assuming all the VLA flux is contained in a milliarcsecond core) corresponding to 54${\,\rm \mu Jy\,beam^{-1}}$ or $6~\times$ VLBI central r.m.s. The region above the blue shaded area represents the AGN dominated regime defined using the selection criteria of \citet{Magliocchetti2018:rl}. The histogram shows the distribution of the radio powers of which peak between $10^{24}$ and $10^{25}$ ${\rm W\,Hz^{-1}}$.}
	\label{Fig:Radio_power_histogram}
\end{figure}

The radio luminosity of an object can be used to determine the probability that the radio emission of a source is dominated by AGN activity or star-formation. The monochromatic 1.5\,GHz radio power can be calculated using:
\begin{equation}
L_{\mathrm{1.5GHz}} = \frac{4\pi d_{L}^2}{(1+z)^{1+\alpha}}S_{\mathrm{obs}},
\end{equation}
where $\alpha$ is the source spectral index, $z$ is the redshift, $d_L$ is the luminosity distance and $S_{\mathrm{obs}}$ is the measured source flux density (ideally, the zero spacing flux density). Integrated flux densities for all 31 sources were derived using the VLA-A array data outlined in Section \ref{SSec:VLA Observations}. Spectral indices for 24/31 sources were derived using the 5.5 GHz integrated flux densities from the VLA 5.5\,GHz survey of \citet{Guidetti:2017wt}. For the seven remaining sources, we used the median spectral index of $-0.56$ from the sources with 5.5\,GHz detections, but we note in passing that these redshift based k-corrections only contribute a small factor to the resulting luminosities. As Figure~\ref{Fig:Radio_power_histogram} shows, we sample a large range of radio powers from $\sim10^{22}\mbox{-}10^{26}~\mathrm{W\,Hz^{-1}}$ which have a median luminosity of $2.7 \times 10^{24}~\mathrm{W\,Hz^{-1}}$. 

As expected, in the low redshift range ($0 < z < 1$), we detect lower luminosity AGN of the order $10^{22}\mbox{-}10^{24}~\mathrm{W\,Hz^{-1}}$ which is typical of objects such as Seyfert galaxies. Higher luminosity sources are not detected at low redshift due to the combination of a low density of high power sources plus a smaller cosmic volume surveyed due to the restricted field-of-view. At higher redshifts, these observations preferentially detect higher luminosity objects of the order $10^{24}\mbox{-}10^{26} \mathrm{W\,Hz^{-1}}$ which is a consequence of the sensitivity (illustrated by the VLBI sensitivity limit plotted in Fig.~\ref{Fig:Radio_power_histogram}). This corresponds to radio-loud AGN systems such as FR-I, radio galaxies and quasars \citep{OwenLedlow:1994fr,2017A&A...602A...6S}. Only three of these sources are lobe-dominated (J123644+621133, J123726+621128 and J123636+615659) with extended morphologies in the lower resolution VLA data while the remaining objects are core dominated, with any jets unresolved or of low luminosity undetectable by the VLA. 

We used the selection criteria of \citet{Magliocchetti2018:rl} to illustrate the radio populations that this survey is probing. Their criteria defines the crossover point, $P_\mathrm{cross}$, as where AGN related emission is dominant over star-formation related emission in a radio-selected population. At $z\leq1.8$, their selection is based upon the radio luminosity functions of \citet{McAlpine2013:RLF}. In this regime, $P_\mathrm{cross}$ approximately scales with redshift as $10^{\log_{10}(P_{0,\mathrm{cross}}) + z}$  where $P_{0,\mathrm{cross}} = 10^{21.7}\,\mathrm{W\,Hz^{-1}\,sr^{-1}}$ which is the crossover point from the local universe. Above $z=1.8$, the radio luminosity function for star forming galaxies drops off rapidly and $P_{\mathrm{cross}}$ is kept constant at $10^{23.5}\,\mathrm{W\,Hz^{-1}\,sr^{-1}}$. This selection criteria was found to keep contaminants from star-forming galaxies to below 10\% at $z>1.8$ \citep[see Appendix of][]{Magliocchetti2018:rl}. 

As Figure~\ref{Fig:Radio_power_histogram} shows, these VLBI observations clearly probe the AGN dominated regime with all radio luminosities at least $3\times P_{\mathrm{cross}}$. For VLBI surveys to truly detect statistically significant samples of hybrid systems with both AGN and star-formation related emission, and provide valuable information regarding feedback, either ultra-wide surveys should be used to probe the tail of star-forming galaxies within the AGN dominated luminosity regime (e.g. \citet{Ruiz2016:rq} investigated radio-quiet AGN using the 2 square degree VLBA survey of COSMOS), or ultra-deep surveys should to used to probe the star-formation dominated luminosity regime which is potentially achievable using the EVN and SKA-VLBI. Nevertheless, we undoubtedly have uncovered hybrid AGN-starburst systems, as we will show in Paper II.

\subsubsection{Brightness temperatures}\label{SSSec:Brightness temperatures}

\begin{figure}[!tb]
	\centering
	\includegraphics[width=\hsize]{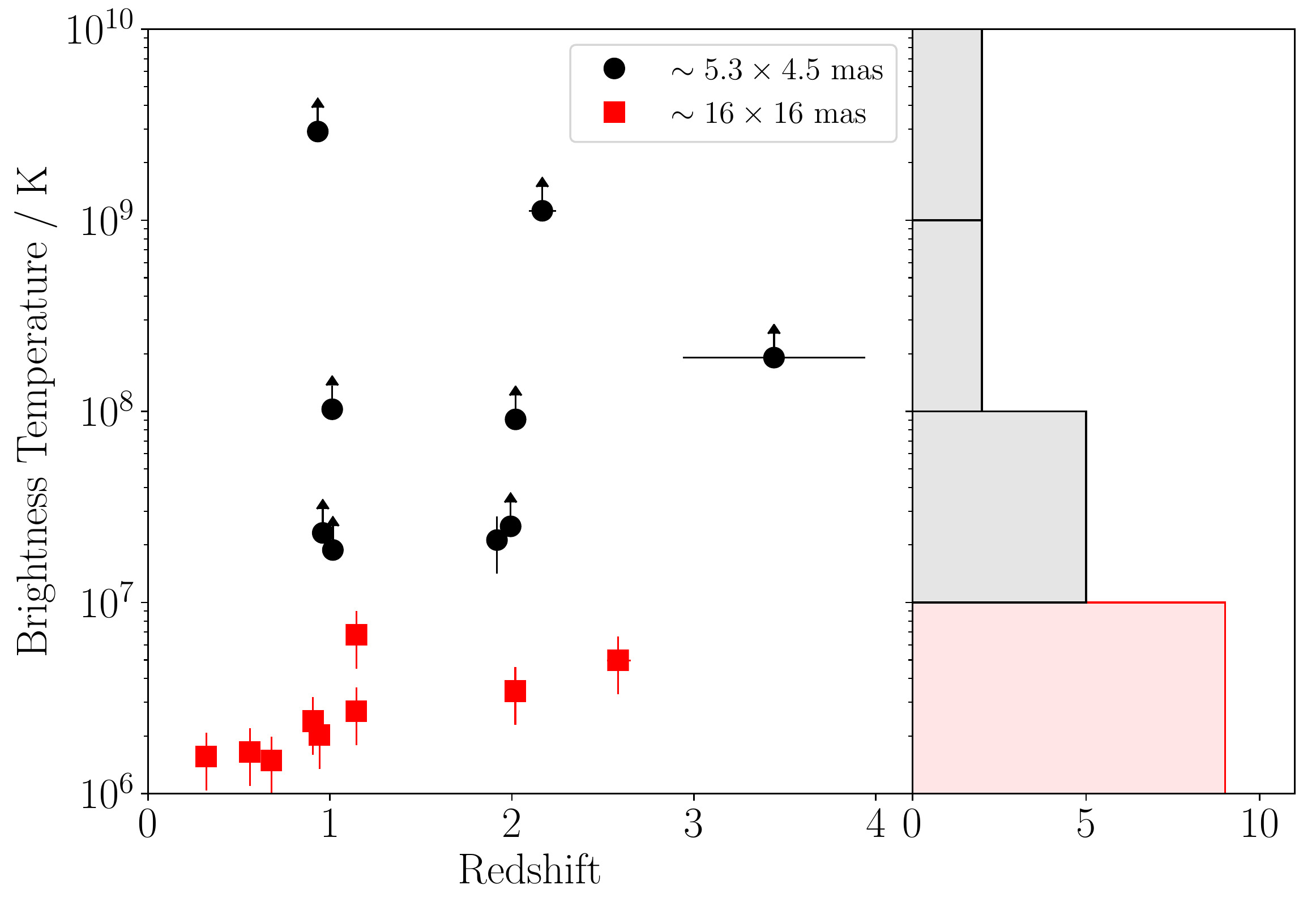}
	\caption{Brightness temperature distribution with respect to redshift. The black circles and red squares correspond to those sources detected with a $\sim5.3\times4.5$ mas restoring beam and those detected with a $\sim 16\times16$ mas beam respectively. Arrows correspond to the lower limits for sources classified as unresolved (that is either $\theta_{\mathrm{maj}}$ or $\theta_{\mathrm{min}}$ is less than the minimum resolvable size). The histogram shows the distribution of brightness temperatures which is colour coded with the markers. The majority of brightness temperatures are between $10^{6}$ and $10^{8}$\,K.}
	\label{Fig:Brightness temperatures}
\end{figure}

Brightness temperatures were calculated for the VLBI detected objects which were primary beam corrected (24/31). For these measurements, we used \texttt{PYBDSF} \citep{Mohan2015:py} to fit an elliptical Gaussian model to each source. We adopted a different selection criterion compared to our detection methodology. In this criterion, referred to as the size measurement detection threshold, sources would be excluded if their S/N ratio were less than ten. This S/N cut-off was chosen because, when using randomly distributed model sources injected onto a noise field from our VLBI data, \texttt{PYBDSF} does not detect all of the injected sources when their $\mathrm{S/N} < 10$. In addition, below this threshold, the variance of fitted sizes is often larger than 20\% of the full-width half-maximum of the psf. 

As we are concerned with only compact emission, we attempted to exclude emission from radio jets or compact star formation by fitting to the uniformly weighted images (with a restoring beam $\sim5.3\times4.5$ mas) if possible. If the S/N of the uniformly weighted image was less than 10, then the naturally weighted images (with restoring beam $\sim16\times16$ mas) would be used to calculate the brightness temperatures. Sources with $S/N < 10$ in both weighting regimes would be excluded completely. Using this selection criteria, 18 sources were selected, 9 using uniform weighting and 9 with natural weighting.

For the calculation, we assume the brightness temperature distribution of a source at redshift $z$ is can be modelled as an elliptical Gaussian radio emission region with major axis $\theta_{\mathrm{maj}}$, minor axis, $\theta_{\mathrm{min}}$, and position angle of the major axis $\phi$. Here in our case, $\theta_{\mathrm{maj}}$, $\theta_{\mathrm{min}}$ and $\phi$ are the deconvolved major and minor axes and their position angle of the deconvolved elliptical Gaussian model. The brightness temperature, $T_{b}$ can then approximated using, 

\begin{equation}\label{equation:brightness temperature}
T_{b} = 1.22 \times 10^{12}(1+z)\left(\frac{S_{\mathrm{\nu}}}{1\mathrm{Jy}}\right) \left(\frac{\nu}{1\mathrm{GHz}}\right)^{-2} \left(\frac{\theta_{\mathrm{maj}}\theta_{\mathrm{min}}}{1\mathrm{mas}^2}\right)^{-1}~\mathrm{K}, 
\end{equation}

\noindent where $S_{\mathrm{\nu}}$ is the observed flux density and $\nu$ is the observing frequency \citep[][]{condon1982temperatures,Ulvestad2005}. In addition, we took into account the resolution limit for both $\theta_{\mathrm{maj}}$ and $\theta_{\mathrm{min}}$ using the prescription described in \citet{lobanov2005}. The following expression for the minimum resolvable size, $\theta_{\mathrm{lim},\psi}$, along each axes of the fitted Gaussian can be calculated using the following equation,

\begin{equation}
\theta_{\mathrm{lim},\psi} = 2^{2-\beta/2}b_\psi \left[\frac{\ln(2)}{\pi}\ln{\left(\frac{\mathrm{S/N}}{\mathrm{S/N}-1}\right)}\right]^{1/2},
\end{equation}

\noindent where $b_\psi$ is the measured FWHM of the psf along the direction of the axis for which the resolution limit is being calculated. S/N is the signal to noise of the image and $\beta$ is a constant that takes into account the weighting of the visibilities ($\beta=0$ for uniform weighting and $\beta=2$ for natural weighting). If $\theta_{\mathrm{maj}}$ or $\theta_{\mathrm{min}}$ were lower than the corresponding minimum resolvable size along each axes, then $\theta_{\mathrm{lim},\psi}$ was used instead to calculate the source frame brightness temperature. A source is classified as unresolved if either axes was below the minimum resolution limit. In this case, the size of the radio emitting region cannot be constrained and only lower limits to the brightness temperature can be derived.

Figure~\ref{Fig:Brightness temperatures} shows the brightness temperature distribution of our VLBI selected sources. Lower limits are derived for those sources which are unresolved or are unresolved in one axis. Sources detected with a 5.3$\times$4.5 mas restoring beam have brightness temperatures of the order $10^7\mbox{-}10^9$ K and, as expected, sources detected with only a 16$\times$16 mas restoring beam have lower brightness temperatures in the range ($10^5\mbox{-}10^6$ K). In both weighting regimes, it is highly unlikely that the radio emission is caused by star-formation related processes as typical star-burst galaxies exhibit brightness temperatures of $<10^5 \mathrm{K}$ \citep{Condon:1991gi}.

Note that the majority of the uniformly weighted sources (8/9) are unresolved, hence emission in these objects come from a compact component. In the naturally weighted sources, all of these are classified as resolved. This is most likely caused by the improved surface brightness sensitivity as a result of the increased weighting of the sensitive, but crucially shorter, central European baselines (especially the Effelsberg to Lovell baseline). As the uniformly weighted images have approximately 1.5 $\times$ the noise of the naturally weighted images, sources detected in natural weighting will most likely have a compact component with higher brightness temperatures, however this component is currently below the size measurement detection threshold (10\,$\sigma$) with uniform weighting. Indeed 6/9 sources do have a compact component in the uniformly weighted images which is above the source detection threshold of 6$\sigma$. The remaining three sources do not have a compact component above $6\sigma$ indicating that some flux may be resolved between the two weighting schemes.

\section{Conclusions}\label{Sec:Conclusions}
We present a catalogue of 31 faint VLBI detected sources in GOODS-N to \textasciitilde$9~\mu {\rm Jy\,beam^{-1}}$ 1$\sigma$ r.m.s. noise levels and radio luminosities of the order $10^{22}~{\rm W\,Hz^{-1}}$. This has substantially increased the number of VLBI detected sources over previous GOODS-N surveys by \citet{chi2013deep} and \citet{garrett2001agn}, providing a valuable addition to the understanding of the AGN content in this well studied field. We also present a primary beam correction scheme developed for the European VLBI Network. This is publicly available and will be constantly updated. 

Additional ancillary information about the radio properties of these objects were derived using VLA\,1.5 and 5\,GHz data. The detected objects have radio luminosities of the order $10^{22}\mbox{-}10^{26}~{\rm W\,Hz^{-1}}$ and brightness temperatures in the range $10^6\mbox{-}10^9~{\rm K}$. The high fraction of compact emission (as defined by the VLBI/VLA flux density ratio) may be hinting at the existence of radio-loud but core-dominated systems at high redshift which may be equivalent to the population of FR0 objects in the local universe \citep{Baldi:2015jt,Baldi2018:em}.

With previous VLBI surveys effectively characterising the $>$50~$\rm \mu Jy$  radio sky, the scheduled 48 hours of remaining observations will enable us to reach limiting r.m.s. sensitivities of $\sim2~{\rm \mu Jy\,beam^{-1}}$, thus providing the next step in analysing the faint radio-selected AGN population in GOODS-N. The final data release will be able to test whether the increasing fraction of star-forming galaxies at these low flux densities are influenced by AGN activity as postulated in lower resolution surveys. This will be presented in a future publication. In addition, the last 24 hours of these observations will include the first ever wide-field VLBI observations using an integrated EVN-eMERLIN array, providing valuable surface brightness sensitivity improvements with the addition of intermediate length baselines. 

\longtab[1]{
	\begin{landscape}
		\begin{longtable}{ccccccccc|cc}
			\caption{1.6\,GHz VLBI and 1.5\,GHz VLA properties of the VLBI detected sources}\label{Table:Source_catalog} \\
			
			\hline\hline
			Source ID & $z$ & $z$ type\tablefootmark{c}/ref  & R.A. (J2000) & Dec. (J2000) & VLBI $P$ & VLBI $I$  & S/N & Beam & VLA $P$ & VLA $I$ \\
			 &  &   &  &  & [$\rm\mu Jy\,beam^{-1}$] & [$\rm \mu Jy$]  &  & [mas$\times$mas (deg)] & [$\rm\mu Jy\,beam^{-1}$] & [$\rm\mu Jy$] \\
			(1) & (2\mbox{-}4) & (5,6) & (7) & (8) & (9,10) & (11,12)  & (13) & (14\mbox{-}16) & (17,18) & (19,20) \\
			\hline
J123555+620902\tablefootmark{a} & 1.8750 & S\tablefootmark{d} & 12:35:55.1267 & +62:09:01.738 & 100.0$\pm$18.2 & 100.0$\pm$18.2 & 7.4 & 16.0$\times$15.2 (87.1) & 165$\pm$17 & 192$\pm$19 \\
J123607+620951 & 0.6380 & S\tablefootmark{d} & 12:36:06.6120 & +62:09:51.159 & 118.0$\pm$22.8 & 118.0$\pm$21.2 & 6.1 & 5.3$\times$4.6 (3.0) & 169$\pm$17 & 205$\pm$21 \\
J123608+621036\tablefootmark{a} & 0.6790 & S\tablefootmark{d} & 12:36:08.1193 & +62:10:35.906 & 122.0$\pm$16.8 & 140.0$\pm$18.2 & 11.1 & 16.0$\times$15.4 (86.8) & 202$\pm$20 & 236$\pm$24 \\
J123618+621541 & 1.9930 & S\tablefootmark{e} & 12:36:17.5546 & +62:15:40.765 & 177.0$\pm$25.0 & 192.0$\pm$26.0 & 10.1 & 5.4$\times$4.6 (10.6) & 226$\pm$23 & 275$\pm$28 \\
J123620+620844 & 1.0164 & S\tablefootmark{d} & 12:36:20.2620 & +62:08:44.268 & 185.0$\pm$25.8 & 185.0$\pm$24.0 & 10.3 & 5.3$\times$4.6 (3.3) & 141$\pm$14 & 156$\pm$16 \\
J123621+621708\tablefootmark{a} & 1.9920 & S\tablefootmark{f} & 12:36:21.2684 & +62:17:08.459 & 96.5$\pm$14.5 & 135.0$\pm$17.3 & 8.9 & 15.6$\times$15.3 (-8.2) & 138$\pm$14 & 190$\pm$19 \\
J123623+620654\tablefootmark{a} & $1.94\substack{+0.12 \\ -0.12}$ & P\tablefootmark{g} & 12:36:22.5086 & +62:06:53.844 & 114.0$\pm$19.0 & 144.0$\pm$21.4 & 8.2 & 16.1$\times$15.3 (86.3) & 222$\pm$22 & 249$\pm$25 \\
J123624+621643 & 1.9180 & S\tablefootmark{e} & 12:36:23.5437 & +62:16:42.746 & 222.0$\pm$28.2 & 383.0$\pm$42.0 & 12.8 & 5.4$\times$4.5 (10.6) & 384$\pm$39 & 411$\pm$41 \\
J123641+621833 & 1.1456 & S\tablefootmark{d} & 12:36:40.5661 & +62:18:33.081 & 141.0$\pm$26.3 & 141.0$\pm$25.7 & 7.5 & 5.3$\times$4.5 (9.4) & 293$\pm$30 & 302$\pm$30 \\
J123642+621331 & 2.0180 & S\tablefootmark{h} & 12:36:42.0899 & +62:13:31.428 & 97.4$\pm$18.0 & 233.0$\pm$27.9 & 6.5 & 5.4$\times$4.5 (12.6) & 432$\pm$44 & 477$\pm$48 \\
J123644+621133 & 1.0128 & S\tablefootmark{d} & 12:36:44.3860 & +62:11:33.170 & 410.0$\pm$44.8 & 411.0$\pm$44.7 & 25.9 & 5.3$\times$4.5 (11.4) & 737$\pm$74 & 1710$\pm$171 \\
J123646+621405 & 0.9610 & S\tablefootmark{d} & 12:36:46.3307 & +62:14:04.692 & 191.0$\pm$24.9 & 192.0$\pm$24.8 & 12.3 & 5.4$\times$4.5 (12.7) & 260$\pm$26 & 280$\pm$28 \\
J123650+620738\tablefootmark{a} & 1.6095 & S\tablefootmark{d} & 12:36:49.6399 & +62:07:37.844 & 77.3$\pm$17.3 & 98.7$\pm$19.9 & 6.5 & 15.4$\times$14.8 (80.9) & 267$\pm$27 & 301$\pm$30 \\
J123653+621444\tablefootmark{a} & 0.3208 & S\tablefootmark{d} & 12:36:52.8827 & +62:14:44.069 & 109.0$\pm$15.1 & 117.0$\pm$15.6 & 11.0 & 14.8$\times$14.7 (9.6) & 188$\pm$19 & 215$\pm$22 \\
J123659+621833 & $2.17\substack{+0.08 \\ -0.07}$ & P\tablefootmark{g} & 12:36:59.3327 & +62:18:32.566 & 2530.0$\pm$328.9 & 4430.0$\pm$572.7 & 88.2 & 5.3$\times$4.5 (8.5) & 4250$\pm$427 & 4640$\pm$464 \\
J123700+620910 & $2.58\substack{+0.07 \\ -0.06}$ & P\tablefootmark{g} & 12:37:00.2460 & +62:09:09.779 & 153.0$\pm$23.4 & 163.0$\pm$24.1 & 9.4 & 5.3$\times$4.5 (8.0) & 272$\pm$27 & 319$\pm$32 \\
J123709+620838 & 0.9070 & S\tablefootmark{l} & 12:37:09.4300 & +62:08:37.587 & 125.0$\pm$21.4 & 127.0$\pm$21.5 & 7.3 & 5.3$\times$4.5 (6.4) & 155$\pm$16 & 163$\pm$16 \\
J123714+621826 & $3.44\tablefootmark{m}$ & P\tablefootmark{i} & 12:37:13.8694 & +62:18:26.301 & 501.0$\pm$56.8 & 629.0$\pm$69.4 & 25.6 & 5.3$\times$4.6 (6.9) & 575$\pm$58 & 637$\pm$64 \\
J123715+620823 & 0.9335 & S\tablefootmark{j} & 12:37:14.9391 & +62:08:23.223 & 2680.0$\pm$272.9 & 2810.0$\pm$284.0 & 103.0 & 5.3$\times$4.6 (5.4) & 1940$\pm$195 & 2090$\pm$209 \\
J123716+621512 & 0.5605 & S\tablefootmark{d} & 12:37:16.3730 & +62:15:12.343 & 125.0$\pm$20.3 & 125.0$\pm$19.7 & 7.9 & 5.4$\times$4.6 (9.9) & 165$\pm$17 & 178$\pm$18 \\
J123717+621733 & 1.1460 & S\tablefootmark{d} & 12:37:16.6800 & +62:17:33.310 & 150.0$\pm$23.8 & 269.0$\pm$32.7 & 8.2 & 5.4$\times$4.6 (7.5) & 308$\pm$31 & 356$\pm$36 \\
J123720+620741\tablefootmark{a} & $0.91\substack{+0.05 \\ -0.03}$ & P\tablefootmark{k} & 12:37:20.0139 & +62:07:41.410 & 94.8$\pm$14.6 & 112.0$\pm$15.8 & 8.8 & 15.9$\times$15.4 (67.2) & 122$\pm$13 & 132$\pm$13 \\
J123721+621130 & $2.02\substack{+0.06 \\ -0.06}$ & P\tablefootmark{g} & 12:37:21.2517 & +62:11:29.961 & 328.0$\pm$38.3 & 364.0$\pm$41.6 & 20.2 & 5.3$\times$4.5 (8.8) & 338$\pm$34 & 385$\pm$39 \\
J123726+621129\tablefootmark{a} & 0.9430 & S\tablefootmark{j} & 12:37:25.9475 & +62:11:28.699 & 124.0$\pm$16.7 & 142.0$\pm$18.2 & 12.2 & 15.4$\times$15.1 (52.6) & 1190$\pm$120 & 5210$\pm$521 \\
\textit{J123649+620439}\tablefootmark{b} & 0.1130 & S\tablefootmark{d} & 12:36:48.9965 & +62:04:38.850 & $>$92.6 & $>$102.0 & 10.5 & 12.5$\times$11.6 (1.2) & 608$\pm$61 & 834$\pm$83 \\
\textit{J123701+622109}\tablefootmark{b} & 0.8001 & S\tablefootmark{d} & 12:37:01.1023 & +62:21:09.623 & $>$111.0 & $>$154.0 & 11.5 & 12.4$\times$11.0 (3.2) & 285$\pm$29 & 390$\pm$39 \\
\textit{J123739+620505}\tablefootmark{b} & $2.99\substack{+0.81 \\ -1.51}$ & P\tablefootmark{k} & 12:37:39.3204 & +62:05:05.489 & $>$154.0 & $>$194.0 & 11.6 & 12.1$\times$10.9 (5.6) & 223$\pm$23 & 258$\pm$26 \\
\textit{J123751+621919}\tablefootmark{b} & $1.20\substack{+0.11 \\ -0.05}$ & P\tablefootmark{k} & 12:37:51.2327 & +62:19:19.012 & $>$111.0 & $>$181.0 & 8.8 & 11.9$\times$10.5 (0.5) & 136$\pm$14 & 155$\pm$16 \\
\textit{J123523+622248}\tablefootmark{b} & $1.42\substack{+0.10 \\ -0.11}$ & P\tablefootmark{k} & 12:35:22.6144 & +62:22:48.028 & $>$92.5 & $>$144.0 & 7.3 & 12.1$\times$10.6 (7.0) & 1260$\pm$126 & 1690$\pm$169 \\
\textit{J123510+622202}\tablefootmark{b} & $2.33\substack{+0.52 \\ -0.24}$ & P\tablefootmark{k} & 12:35:10.2698 & +62:22:02.067 & $>$88.9 & $>$91.4 & 7.9 & 12.1$\times$10.6 (7.0) & 931$\pm$94 & 1280$\pm$128 \\
\textit{J123656+615659}\tablefootmark{b} & $0.39\substack{+0.05 \\ -0.04}$ & P\tablefootmark{k} & 12:36:55.8230 & +61:56:58.917 & $>$518.0 & $>$528.0 & 12.7 & 9.6$\times$9.0 (39.0) & 3590$\pm$361 & 26700$\pm$2670 \\
\hline
			
			\hline\end{longtable}
		\tablefoot{z: redshift, R.A.: Right Ascension (J2000), Dec.: Declination (J2000), VLBI $P$: VLBI peak brightness ($\mathrm{\mu Jy\,beam^{-1}}$), VLBI $I$: VLBI integrated flux density ($\mathrm{\mu Jy}$), N: noise ($\mathrm{\mu Jy\,beam^{-1}}$), S/N: signal-to-noise, Beam: restoring beam in milliarcseconds and beam angle in degrees (major axis $\times$ minor axis (beam angle)), VLA $P$: VLA 1.5\,GHz peak brightness,  VLA $I$: VLA 1.5\,GHz integrated flux densities. Italiscised source IDs correspond to sources with no-primary beam correction applied. The row of numbers below the column titles correspond to the columns in the machine-readable table that accompanies this paper.\\
			\tablefoottext{a}{Sources detected using naturally weighted taper (\texttt{UVWTFN=`NA'} in AIPS task \texttt{IMAGR})} \tablefoottext{b}{Not primary beam corrected.}
			\tablefoottext{c}{S: spectroscopic redshift, P: photometric redshift.} Redshift references: \tablefoottext{d}{\citet{Barger_specz_2008}}, \tablefoottext{e}{\citet{Smail2004:sz}},\tablefoottext{f}{\citet{Chapman:2005wh}},\tablefoottext{g}{\citet{Skelton_HST3D_2014}},\tablefoottext{h}{\citet{Murphy:2017ja}},\tablefoottext{i}{\citet{Cowie2017z}},\tablefoottext{j}{Cowie priv. comm.},\tablefoottext{k}{\citet{Yang_photz_2014}},\tablefoottext{l}{\citet{Cowie2004z}}. \tablefoottext{m}{Unknown photometric error, conservatively set to $\pm0.5$ in calculations of derived properties}}
\end{landscape}}

\longtab[1]{
		\begin{longtable}{cccccc}
			\caption{Derived VLA \& VLBI radio properties of the 31 GOODS-N AGN.}\label{Table:Derived_properties}  \\
			
			\hline\hline
			Source ID &  $\alpha$& $L_\mathrm{{1.5GHz}}$ & $T_b$ & Angular sizes & Linear sizes \\
			 &  & [$\rm W\,Hz^{-1}$] & [$\rm K$] & [mas] & [parsec] \\
			(1) & (21) & (22,23) & (24\mbox{-}26) & (27\mbox{-}30) & (31\mbox{-}34) \\
			\hline
J123555+620902 & - & $(2.7\pm 0.3) \times 10^{{24}}$ & - & - & - \\
J123607+620951 & $-1.02$ & $(3.1\pm 0.3) \times 10^{{23}}$ & - & - & - \\
J123608+621036 & $-0.46$ & $(3.3\pm 0.4) \times 10^{{23}}$ & $ \mathit{1 \times 10^{6}}$ & 11.1$\times$6.3 & 80.8$\times$45.8 \\
J123618+621541 & $-0.62$ & $(4.5\pm 0.4) \times 10^{{24}}$ & $>3 \times 10^{7}$ & 3.7$\times$$<$2.8 & 31.5$\times$$<$23.8 \\
J123620+620844 & $-0.28$ & $(4.9\pm 0.6) \times 10^{{23}}$ & $>2 \times 10^{7}$ & $<$3.2$\times$$<$2.8 & $<$26.3$\times$$<$22.9 \\
J123621+621708 & $-0.78$ & $(3.3\pm 0.3) \times 10^{{24}}$ & - & - & - \\
J123623+620654 & $0.06$ & $(2.0\pm 0.7) \times 10^{{24}}$ & - & - & - \\
J123624+621643 & $-0.52$ & $(6.3\pm 0.7) \times 10^{{24}}$ & $2 \times 10^{7}$ & 5.9$\times$4.0 & 50.8$\times$34.2 \\
J123641+621833 & $-0.94$ & $(2.2\pm 0.2) \times 10^{{24}}$ & $ \mathit{3 \times 10^{6}}$ & $<$12.3$\times$5.0 & 104.4$\times$42.6 \\
J123642+621331 & $-1.05$ & $(1.4\pm 0.1) \times 10^{{25}}$ & $ \mathit{3 \times 10^{6}}$ & 12.1$\times$8.5 & 103.4$\times$73.2 \\
J123644+621133 & $-0.56$ & $(3.1\pm 0.3) \times 10^{{24}}$ & $>1 \times 10^{8}$ & 2.1$\times$$<$1.7 & 17.6$\times$$<$13.9 \\
J123646+621405 & $-0.40$ & $(8.7\pm1.0) \times 10^{{23}}$ & $>2 \times 10^{7}$ & $<$2.9$\times$$<$2.5 & $<$23.9$\times$$<$20.1 \\
J123650+620738 & $-0.56$ & $(3.1\pm 0.3) \times 10^{{24}}$ & - & - & - \\
J123653+621444 & $-0.11$ & $(5.3\pm 0.7) \times 10^{{22}}$ & $ \mathit{2 \times 10^{6}}$ & 9.2$\times$4.8 & 44.1$\times$23.0 \\
J123659+621833 & $-1.19$ & $(2.0\pm 0.1) \times 10^{{26}}$ & $>1 \times 10^{9}$ & 6.2$\times$$<$0.9 & 52.2$\times$$<$7.7 \\
J123700+620910 & $-0.89$ & $(1.3\pm 0.1) \times 10^{{25}}$ & $ \mathit{5 \times 10^{6}}$ & $<$9.5$\times$7.2 & 78.3$\times$59.1 \\
J123709+620838 & $0.15$ & $(3.2\pm 0.5) \times 10^{{23}}$ & $ \mathit{2 \times 10^{6}}$ & 7.8$\times$6.1 & 63.0$\times$49.1 \\
J123714+621826 & $-0.66$ & $(3.9\pm1.2) \times 10^{{25}}$ & $>2 \times 10^{8}$ & 3.8$\times$$<$1.7 & 28.5$\times$$<$12.9 \\
J123715+620823 & $-0.04$ & $(4.8\pm 0.7) \times 10^{{24}}$ & $>3 \times 10^{9}$ & $<$1.0$\times$$<$0.8 & $<$7.9$\times$$<$6.9 \\
J123716+621512 & $-0.19$ & $(1.5\pm 0.2) \times 10^{{23}}$ & $ \mathit{2 \times 10^{6}}$ & 10.4$\times$6.5 & 69.1$\times$43.4 \\
J123717+621733 & $-0.89$ & $(2.2\pm 0.2) \times 10^{{24}}$ & $ \mathit{7 \times 10^{6}}$ & 6.8$\times$5.1 & 57.6$\times$43.2 \\
J123720+620741 & $-0.28$ & $(3.4\pm 0.6) \times 10^{{23}}$ & - & - & - \\
J123721+621130 & $0.01$ & $(3.5\pm 0.8) \times 10^{{24}}$ & $>9 \times 10^{7}$ & 2.8$\times$$<$1.9 & 24.0$\times$$<$16.5 \\
J123726+621129 & $-1.23$ & $(6.6\pm 0.4) \times 10^{{24}}$ & $ \mathit{2 \times 10^{6}}$ & 8.7$\times$6.9 & 71.0$\times$56.1 \\
\textit{J123649+620439} & - & $(2.0\pm 0.4) \times 10^{{22}}$ & - & 8.3$\times$6.0 & 17.7$\times$12.7 \\
\textit{J123701+622109} & - & $(7.0\pm 0.7) \times 10^{{23}}$ & - & 9.4$\times$7.3 & 72.8$\times$56.8 \\
\textit{J123739+620505} & - & $(9.7\pm8.9) \times 10^{{24}}$ & - & 8.6$\times$7.2 & 67.9$\times$56.7 \\
\textit{J123751+621919} & - & $(8.2\pm1.6) \times 10^{{23}}$ & - & - & - \\
\textit{J123523+622248} & - & $(1.1\pm 0.2) \times 10^{{25}}$ & - & - & - \\
\textit{J123510+622202} & - & $(2.4\pm 1.0) \times 10^{{25}}$ & - & - & - \\
\textit{J123656+615659} & - & $(1.7\pm 0.5) \times 10^{{24}}$ & - & 7.3$\times$$<$2.4 & 39.7$\times$$<$13.2 \\
			\hline
		\end{longtable}
		\tablefoot{$\alpha$: 1.5GHz\mbox{-}5.5\,GHz spectral index, $L_\mathrm{{1.5GHz}}$: monochromatic 1.5\,GHz radio luminosity, $T_b$: brightness temperature (italicised indicates that natural weighting was used to derive $T_b$),  Angular size: projected angular size using elliptical Gaussian fitting, Linear size: projected linear size in parsecs. Italiscised source IDs correspond to sources with no-primary beam correction applied. Row of numbers below the column titles correspond to the columns in the machine-readable table that accompanies this paper.}
}

\begin{acknowledgements}\label{Sec:Acknowledgements}
We thank the anonymous referee for their comments which helped improve this manuscript. 
The authors gratefully acknowledge Len Cowie and Amy Barger for allowing us to use unpublished data and for their useful comments. We would also like to thank P. Padovani and R. Windhorst for their useful discussions. \\
This research made use of Astropy, a community-developed core Python package for Astronomy \citep{Astropy,Astropy2018}. \\
The research leading to these results has received funding from the European Commission Seventh Framework Programme (FP/2007-2013) under grant agreement No 283393 (RadioNet3).\\
The European VLBI Network is a joint facility of European, Chinese, South African, and other radio astronomy institutes funded by their national research councils. \\
e-MERLIN is a National Facility operated by the University of Manchester at Jodrell Bank Observatory on behalf of STFC.\\
The National Radio Astronomy Observatory is a facility of the National Science Foundation operated under cooperative agreement by Associated Universities, Inc.\\
\end{acknowledgements}

%
\bibliographystyle{aa} 
\bibliography{Nowhere_to_Hide} 
%

\begin{appendix}
\section{Astrometry of J1234+619}\label{appendix}
\begin{figure}[!tb]
	\centering
	\includegraphics[width=\hsize]{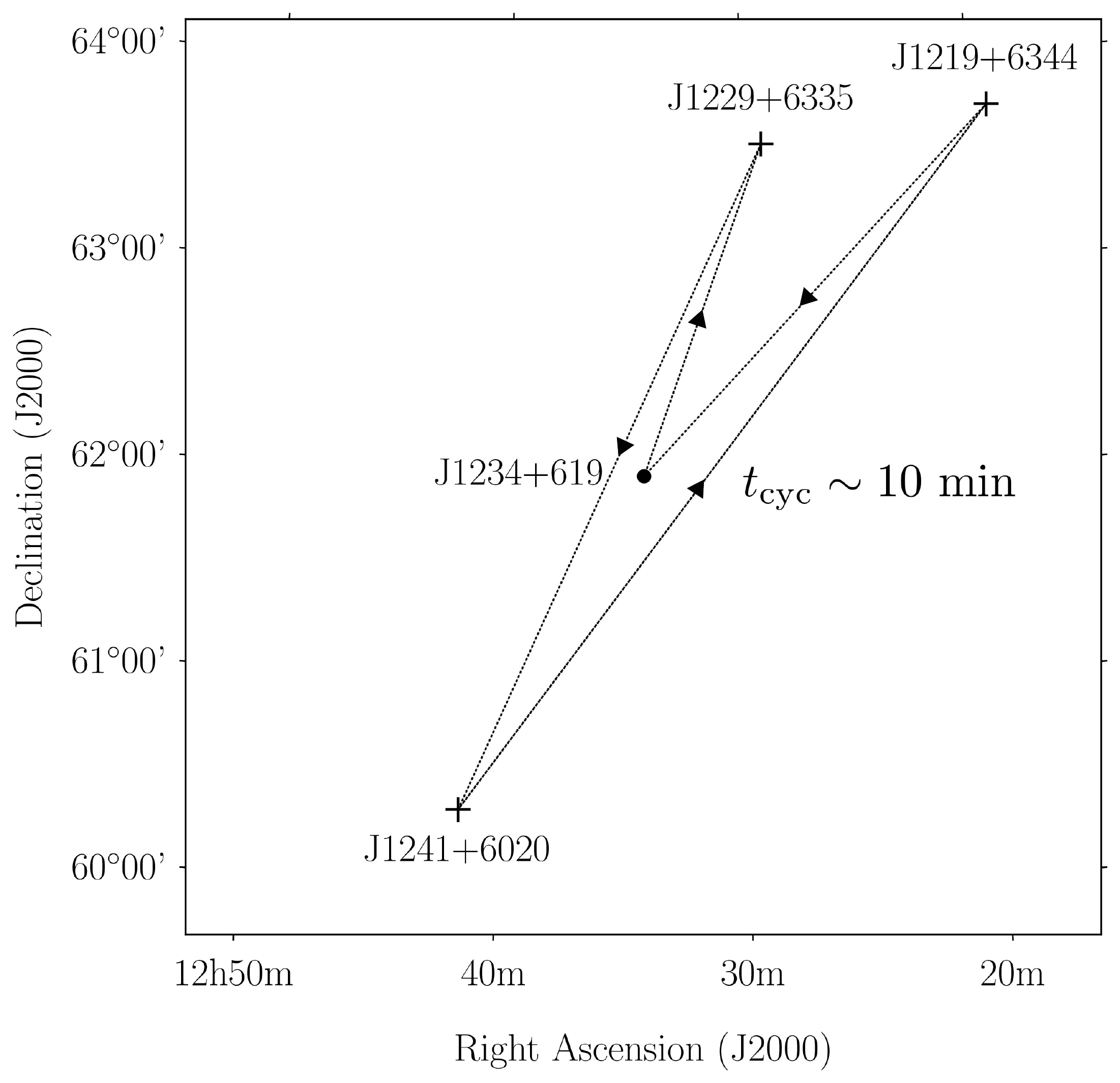}
	\caption{Sky distribution of sources observed with e-MERLIN at 5 GHz. Table~\ref{Table:phs_cals} lists the source coordinates and associated astrometric errors. The dashed arrows correspond to the source switching order of the observations. 
	}
	\label{Fig:Source positions_astrometry}
\end{figure}

We acquired 12 hours of e-MERLIN time at 5\,GHz in order to independently confirm the position of J1234+619. In order to effectively characterise changes in the delay path induced by tropospheric variations, we implemented a multi-source phase referencing strategy using three VLBA calibrators surrounding J1234+619. These sources have astrometric accuracies $<2.6~{\rm mas}$. Whilst not as sophisticated as astrometric techniques such as MultiView calibration \citep{Rioja:2017kb}, multi-source phase referencing has been used to obtain astrometric precisions of $\sim 20 {\rm \mu as}$ with the VLBA \citep{Fomalont:2002wu}. We observed three calibrators and then the target with a total cycle time, $t_{\rm cyc}$, of $\sim 10~{\rm min}$ (summarised in Figure~\ref{Fig:Source positions_astrometry} and Table~\ref{Table:phs_cals}). The use of multiple phase calibrators in different relative positions allows the spatial dependence of the phase to be resolved and a short cycle time is desired in order to map the atmosphere-induced phase variations across all sources more frequently.

In order to estimate the uncertainties in the astrometric measurements, we consider the approach of \citet{Rioja:2017kb}. Errors arise from three main sources: (1) Thermal noise which originates from the uncertainty in the position measurement due to random noise. This is characterised as $\sigma_t \sim \theta_B / 2(\rm S/N)$ where $\theta_B$ is the restoring beam and S/N is the signal-to-noise. The target field has a S/N of 182, hence the thermal noise error is approximately 0.11 mas (2) Fractional flux recovery. This is the difference between the peak brightness in the phase referenced maps and the self-calibrated maps. (3) Accuracy. The observed phase reference sources have astrometric errors $<2.6 {\rm mas}$ with 2/3 having errors $<{\rm0.5mas}$. Positional errors associated with effects such as the core-shift effect, the use of phase delays over geodetic group delays \citep{Porcas:2009ka} and differential structure blending effects between the EVN VLBI observations at 1.6 GHz and these observations at 5 GHz is expected to be a few milliarcseconds. In total, errors on these observations is expected to be dominated by unknown effects such as core-jet blending. We therefore assign a conservative estimate of $\leq$ 5 mas astrometric accuracy. 

With the phase referencing from the three phase calibrators complete, these solutions were transferred to J1234+619 and positions on the resultant image were measured using AIPS task JMFIT. The results of these observations are presented in Table~\ref{Table:phs_cal_results}. We find that the eMERLIN 5\,GHz positions of J1234+619 are more consistent with those derived as part of these new EVN observations, rather than than the position of \citet{chi2013deep}. However, due to the unknown errors, such as core-jet blending and structure blending associated with the differing resolutions, we will continue to use the positions derived in these new VLBI observations, as they are more consistent with the 5\,GHz eMERLIN positions. 

\begin{table}
	\caption{Phase calibrators used in the 5GHz eMERLIN observations}             
	\label{Table:phs_cals}      
	\centering  
	\begin{tabular}{lllll}
		\hline\hline 
		IAU Name & R.A. (J2000) & Dec. (J2000) & $\Delta$RA & $\Delta$Dec  \\
		& & & [mas] & [mas] \\
		\hline
		J1229+6335 & 12:29:06.0260 & 63:35:00.980 & 1.80 & 2.58 \\
		J1241+6020 & 12:41:29.5906 & 60:20:41.322 & 0.27 & 0.29 \\
		J1219+6344 & 12:19:10.5831 & 63:44:10.718 & 0.34 & 0.48 \\
		\hline
	\end{tabular} 
	\tablefoot{Phase calibrators and their associated astrometric uncertainties. Positions acquired from VLBA calibrator list\footnote{\url{http://www.vlba.nrao.edu/astro/calib/}} \citep{Fey:2015jc}. Abbreviations: RA: Right Ascension (J2000), Dec: Declination (J2000), $\Delta$RA: error in RA (mas), $\Delta$Dec: error in Dec (mas).}
\end{table} 
\begin{table}
	\caption{Positions and relative offsets of J1234+619}             
	\label{Table:phs_cal_results}      
	\centering 
	\begin{tabular}{llll}
		\hline\hline
		& R.A. (J2000) & Dec. (J2000) \\
		\hline
		EVN &  12:34:11.7413(57) & +61:58:32.478(07) \\
		C13 &  12:34:11.7430 & +61:58:32.480 \\
		eM5  & 12:34:11.7417(10) & +61:58:32.477(68) \\
		\hline
	\end{tabular}
	\begin{tabular}{lllll}
		\hline\hline
		& $\Delta\theta$ & $\Delta$R.A. (J2000) & $\Delta$Dec. (J2000) \\
		& [mas] & [mas] & [mas] \\
		\hline
		C13-EVN & 11.74 & 11.58 & 1.93 \\
		C13-eM5 & 9.39 & 9.09 & 2.32 \\
		EVN-eM5 &  2.52 & 2.48 & 0.39 \\
		\hline
	\end{tabular}
	\tablefoot{{\it Top Panel:} Positions of J1234+619 from \citet{chi2013deep} (C13), this paper (EVN), and eMERLIN 5GHz (eM5). {\it Bottom Panel:} Angular offsets between the positions derived. Abbreviations: RA: Right Ascension (J2000), Dec: Declination (J2000), $\Delta$RA: error in RA (mas), $\Delta$Dec: error in Dec (mas), $\Delta \theta$: angular separation.}
\end{table}
\end{appendix}

\end{document}